\documentclass[12pt,a4paper]{article}
\usepackage{times}
\usepackage{newtxmath}
\usepackage[T1]{fontenc}
\usepackage{a4wide}
\usepackage{textcomp,makeidx}
\usepackage{color,stabular}
\usepackage{listings}
\usepackage{hyperref}
\hypersetup{colorlinks=true,linkcolor=blue,urlcolor=blue}
\newcommand\ie{\textit{i.e.,\ }}
\newcommand\eg{\textit{e.g.,\ }}
\newcommand\lhs{\textit{l.h.s.}}
\newcommand\rhs{\textit{r.h.s.}}
\newcommand\FRI{{FRI}}
\newcommand\CPL{{CPL}}
\newcommand\CPLlogo{{\rm C\hspace{-0.1em}\protect\raisebox{0.35ex}{P}\hspace{-0.41em}L}}
\definecolor{codebg}{gray}{0.9}
\newsavebox{\codebox}
\newenvironment{code}
{\medskip\par\noindent
\begin{lrbox}{\codebox}\begin{minipage}{0.983\textwidth}\small\tt
\hangindent=2em
\raggedright\obeyspaces\obeylines}
{\end{minipage}\end{lrbox}\colorbox{codebg}{\usebox{\codebox}}\medskip\par\noindent\ignorespacesafterend}

\newcommand\cod[1]{\hspace{0pt}\linebreak[0]\colorbox{codebg}{\small\tt #1}}

\newcommand\auxrefcode[1]{
\noindent\hyperref[#1-code]{\normalfont $\rightarrow$code}
}

\newcommand\auxcoderef[1]{
\hyperref[#1]{\normalfont description$\leftarrow$}
}
\title{Introducing \CPLlogo}
\author{Paolo Luchini\\
\texttt{luchini@CPLcode.net}}

\lstdefinelanguage{cpl}
{sensitive=true,
texcl=true,
comment=[l]{!!}
}
\makeindex

\begin{document}
\maketitle
\begin{abstract}
CPL here stands for a computer programming language conceived and developed by the author since 1993, but published for the first time in 2020. It was born as a Compiled Programming Language, designed together with its compiler and therefore suitable for computationally intensive numerical applications, although some years later an interpreter was also provided for interactive usage. CPL's distinctive features are Concealed Pointer Lookup, the ability to implicitly dereference pointers based on the type of operands involved, Consistent Procedure Linkage, the enforcement of function prototypes without dedicated header or interface files, and Coactive Parameter Lists, the ability to overload function names which are then distinguished by the type of their parameters and/or parameter separators. Perhaps even more distinctly, CPL's syntax can be extended on the fly by the program being compiled; library modules tap this feature to seamlessly add real and complex matrix operations, graphics, parallel-computing extensions, and symbolic differentiation. The CPL coding software is available for free download at \url{http://CPLcode.net}.
\end{abstract}
\tableofcontents
\label{CPLlang}

\section{Introduction}

``CPL'' is a versatile acronym with a multitude of meanings, among which more than one published programming language. Here it will be used to denote a Compiled Programming Language Created by Paolo Luchini since 1993, but published for the first time in 2020, endowed with Concealed Pointer Lookup, Consistent Procedure Linkage and Coactive Parameter Lists. The CPL development software, comprising both a source-to-source compiler and an interpreter, has been privately available to a small number of active scientific researchers (in the author's principal subject area which happens to be Fluid Dynamics and Turbulence) for almost three decades; their field experience contributed to honing the details, and their enduring predilection for this over more standard programming languages convinced the author to now make CPL publicly available for free download at \url{http://CPLcode.net}. The present text constitutes its language definition and reference manual.

The first question everybody wil want answered, before actually being enticed to read on, is ``Why yet another programming language?'' Therefore the exposition will start with the features that distinguish \CPL\ most, and that were at the same time the driving force behind its development. It will end with an alphabetical list of CPL keywords [\hyperlink{Keyword_index}{Keyword index}] and a partial list of scientific papers that already used CPL for their numerical programming needs [\hyperlink{References}{References}].

Another, mostly aesthetic but distinctive, feature of CPL is the availability of all three kinds of round, square and curly brackets for grouping arithmetic (or other) expressions, just as you were taught in school (see \S\ref{Parentheses}). While this might not be a game-changing feature, from a human-interface viewpoint to alternate brackets so much eases reading and proofreading nested parentheses, and it increases the chances of automatic error catching at the same time.
The provided text editor can automatically complete parentheses and alternate bracket types for you.

\subsection{Concealed Pointer Lookup\label{Concealed_Pointer_Lookup}\index{Implicit type conversion}}
Pointers are the connecting link between high-level and machine language. Pointer operations closely reproduce the address manipulations that the processor and its machine language are so fast at doing, but also the pitfalls and hard-to-chase errors that address manipulation brings along. Thus the challenge for a high-level language is to provide as much as possible of this flexibility and speed while at the same time making as many errors as possible automatically highlighted. The most frequent usage of pointers is in one of three roles: as links in linked structures, as cursors for array traversal, and as subroutine parameters passed by reference. The very first target of \CPL\ at its birth, was to strike a middle ground between the opposite extreme approaches that the then-mainstream FORTRAN 77 and Kernighan\&Ritchie C languages had adopted in their pointer handling.

The FORTRAN-77 approach is to drastically hide all evil of pointers. Arrays are only traversed by integer indices, linked structures have to be mimicked through arrays, and all subroutine parameters are implicitly passed by reference, effectively as pointers but without making this apparent in the language.

K\&R (but also ANSI) C's approach is just the opposite: to give pointers the maximum of flexibility but also of visibility with their omnipresent \cod{\&} and \cod{*} prefix operators, which respectively take the address of a given variable and dereference a pointer to the value it points to.

Both approaches have their weaknesses, one because it requires convoluted (and therefore error prone) constructions in the places where pointers would be appropriate, the other because it forces the programmer to reckon with pointers even where they could be handled automatically, and allows so much flexibility that legal and illegal usages often become hardly distinguishable.

Thus, the first design requirement of \CPL\ was that it should have \emph{Concealed Pointer Lookup}, the capability to handle pointers where needed, but also to conceal the address-extraction and dereferencing operations where the compiler could automatically infer from the context that they are intended. The required information is, in fact, most of the time already available through type-consistency checking. When a variable is passed as a parameter to a function or an operator, the compiler typically knows whether its value or its address is expected, and can perform the required conversion without an explicit address-taking or dereferencing symbol.

Type consistency is enforced through a chain of implicit type conversions.  Allowed conversions are automatically walked through in an attempt to build a consistent expression. If this turns out impossible, a compilation error results. The basic rule is that a type can be implicitly converted to one other type only, and that there cannot be closed type-conversion loops. Just as \cod{INTEGER}s are implicitly converted to \cod{REAL}s, \cod{POINTER}s are
implicitly converted to the value they point to. Owing to this Concealed
Pointer Lookup mechanism, it is rarely necessary to explicitly dereference pointers. (A notable exception is the equals \cod{=} sign, either in its use to denote an assignment or a comparison, which is so heavily overloaded that dereferencing must be explicitly specified in order to distinguish, \eg the assignment of an address to a \cod{POINTER} variable from the assignment of a value to the field it points to. The way to do so will be presented in \S\ref{Assignment}.)

\subsection{Consistent Procedure Linkage}

FORTRAN-77 had no consistency check across separately compiled files. It was the programmer's full responsibility to ensure that subroutines are called consistently with their definitions. The C language also enforced, and still enforces, no intrinsic consistency check outside the bounds of a single compilation unit, but its universally adopted programming style requires the use of headers, files of function prototypes that get included in such a way that intra-file checking acquires an inter-file significance. Fortran-90 follows a similar approach with its \cod{INTERFACE}.
Modula-2 made this two-file structure mandatory, by requiring so called \cod{DEFINITION} and \cod{IMPLEMENTATION} modules for each program segment.

Consistent procedure linkage is an overly useful feature, as anybody appreciates who has progressed from FORTRAN-77 to a language that has it. It is also a factor in the popularity of interpreted languages, which handle inter-file calls at run time. However, keeping separate definition and implementation files for every program segment is a bit inconvenient, because essentially the same information has to be written twice (and manually kept consistent across versions). So, why not to have one file achieve both purposes?

\CPL\ achieves Consistent Procedure Linkage with one file per module only. The trick is in the \cod{USE} statement detailed in \S\ref{USE}, which seamlessly integrates some project building (\texttt{make}) features in the compiler. \cod{USE} preventatively looks for a possibly present .o object file having the same base name as the \texttt{.cpl} file, and decides based on its modification date. If the object file does not exist or is older, the \cod{USE}d \texttt{.cpl} file is separately compiled. If a newer \texttt{.o} file already exists, no new compilation is started and the \texttt{.cpl} file is read as though it was a header, this meaning that outermost-block declarations are interpreted and assimilated in the dictionary so they become effective for error checking purposes, but all inner blocks, including subroutine bodies, are just skipped over. This scheme still allows one the option to distribute header files that do not reveal the program's source code, if really wanted, because empty subroutine bodies are acceptable once a suitable object file exists.

\subsection{Coactive Parameter Lists\label{Coactive_Parameter_Lists}\index{Coactive Parameter Lists}}

\CPL\ has Coactive Parameter Lists, meaning that the type definition of a
function's parameter list coacts with its name in identifying the function or
subroutine. In other words, functions of the same name and different parameter
types are allowed and considered distinct by the language, a feature alternately
known as function overloading. In addition, not just commas but any
non-ambiguous single character or character sequence (except semicolon, spaces or comments, newline and closing brackets) may appear as the separator between actual parameters of a function call. In formal-parameter declarations, non-alphanumeric separators can be
specified verbatim, whereas separators that could be mistaken for identifiers
are to be enclosed in (either single or double) quotes. In subroutine and
function calls the quotes are omitted, and the compiler enforces conforming
separators. Whereas commas will still be the commonest choice for short
parameter lists, the adoption of remindful separators can make long parameter
lists both more understandable and more thoroughly checked by the compiler.

Formal parameter declarations otherwise follow the same syntax as general
variable declarations, and are distinguished both by their position within the
parenthesis following the subroutine or function name and by separator
matching. Optional parameters recognized by name are also available (see \S\ref{OPTIONAL}). Round, square or curly brackets can be used, just as everywhere else
(see \S\ref{Parentheses}). However an important difference exists between parameters and variables: whereas general
variables are by default \cod{VARIABLE} (see \S\ref{VARIABLE}), i.e. can be assigned a value
multiple times, and must be explicitly declared \cod{CONSTANT} (see \S\ref{CONSTANT}) to become
inalterable, formal parameters are by default \cod{CONSTANT} and only become \cod{VARIABLE}
if explicitly declared so. Parameters can be \cod{OPTIONAL} (see \S\ref{OPTIONAL}), and be specified by name rather than by position.

A subroutine returns at the END of the subroutine block or at the EXIT (see \S\ref{EXIT}) 
statement. Within functions a variable named RESULT (see \S\ref{RESULT}) is implicitly declared with
the type of the function's result, and can be used just as any other local
variable of that type within the body of the function; the function returns the
value of RESULT at the time of exit.

The declared subroutine is visible from the point of its declaration to the
end of the enclosing block, just as all other declarations. This includes the
subroutine body itself, and recursive calls are thus automatically allowed.

A Coactive Parameter List means that function names can be overloaded, that is used multiple times and recognized on the basis of their parameter lists, either through different parameter types or through different parameter separators. This achieves a purpose similar to object programming, in the sense that one function name may semantically express several similar actions, each to be performed on a different object or set of objects. Dynamically typed objects are also cared for (see \S\ref{DYNAMIC}). In fact, this feature is ampler than typical object programming because the type of all arguments, not just of the first, is significant.  

\subsection{Extensible syntax\label{FRI_SECTION}\index{Extensible syntax}}

One design decision that was instrumental to the gradual development of \CPL\ while it was already being used, is the concept of a compiler-interpreter.
The design of most compilers passes through a formal description of the         
language that is compiled by a parser generator like \texttt{yacc} into a C program. At
the heart of CPL's extensibility there is a different structure, where a formal
description of the language is never compiled to generate the compiler, but
rather interpreted each time a compilation is run. The engine that does this
is named \texttt{fri}, the Formal Rule Interpreter. If \texttt{yacc} is a compiler-compiler, \texttt{fri} may be called a compiler-interpreter.

The \CPL\ compiler is internally driven by a set of formal rules written in the \FRI\ declarative language. Such rules are stored in the dictionary of the \texttt{fri} compiler-interpreter during every compilation, and can be dynamically modified by suitable directives.
\FRI\ language excerpts may be inserted at any point of a \CPL\ program using the syntax

\begin{code}
FRI SECTION

\emph{some FRI code}

END FRI SECTION
\end{code}

The \cod{FRI SECTION} declaration has the effect of dynamically adding new constructions and for
all purposes extending the syntax of the language. This is not a general user's easy feat, as the new rules have to be consistent with already present ones and the FRI language is not covered in the present guide, but this is how the extensions contained in the pre-packaged \CPL\ libraries are implemented. Like
everything else in \CPL, the new extensions become available from the point where they occur through the end of the enclosing block (which can be the whole program if \cod{FRI SECTION} appears at the outermost level).

\subsection{Libraries}
As should by now be clear, since the first instant the primary goal of \CPL's design was extensibility. Many extended features are therefore available as library modules that can be loaded on demand, and such features become seamlessly integrated in the syntax of the language. Among them are
\begin{enumerate}
\item Complex numbers
\item Fast Fourier transforms
\item Matrix algebra
\item Parallel computing
\item Plotting (gnuplot interface)
\item Runtime bounds checking
\item Symbolic manipulation
\item Interactive execution
\end{enumerate}
The first few of these are self-explanatory, 
but run time bounds checking, symbolic manipulation and interactive execution may deserve some words of explanation.
                                                                               
\subsubsection{Run time bounds checking}\index{Run time bounds checking}
The \texttt{rtchecks.cpl} library does not add any new subroutines or instructions, but  
modifies already existing ones in such a way as to provide run-time bounds      
checking. The impact on performance is considerable; therefore \texttt{rtchecks} should  
only (and always) be used during development and later disabled once the        
program is deemed to be reliable. In particular:                                
                                                                                
All array indexing and subarray extraction operations are bounds-checked,       
pointers are initialized to \cod{NULL} and \cod{NULL}-pointer dereferencing is trapped,     
all error messages are labelled so as to display the position in the source
code where the error occurred.                                                  

\subsubsection{Symbolic manipulation}\index{Symbolic manipulation}
Symbolic manipulation is the feat of processing an algebraic expression in terms of its symbols, rather than numerical value, the most eminent incarnation being symbolic differentiation; while famous symbolic manipulation programs like \texttt{MACSYMA} or \texttt{Maple} have become part of the history of computing, this is not an idea that is usually associated with compilers. Nevertheless, one of the interesting usages of symbolic manipulation is to generate code that is subsequently fed to a compiler, a technique known as Automatic Differentiation (AD). A compiler-interpreter, like the \CPL\ compiler is, can make automatic differentiation happen at compile time. The key idea is to differentiate program expressions on the spot before submitting them to actual compilation. The \texttt{symbolic.cpl} library achieves this feat by suitably extending the concept of an alias, or deferred assignment (\S\ref{Deferred_assignment}), that is a symbol representing a formula that is not to be evaluated at the position in the program where its definition appears but at the later position where its value is used. After such an alias is defined, it comes natural to apply symbolic operations to it, which are what \texttt{symbolic.cpl} enables; such operations generate another alias as their result, without producing any immediate code; the new aliases, after possibly being subjected to more symbolic operations, will then be compiled into executable code at the position where each of them is eventually evaluated.

\subsubsection{Interactive execution}\index{Interactive execution}
Interactive execution is what an interpreter rather than a compiler provides. Being \texttt{fri} a compiler-interpreter it actually provides both, and the \texttt{icpl} library (constructed in such a way that, under a unix system, the standalone \texttt{icpl} command launches \texttt{fri} with the library in it) adds such a capability and the few language extensions necessary to make it comfortable. It achieves this purpose through a pipeline of \FRI\ rules that in real time feeds every instruction typed on the command line to the already existing \CPL-to-C compiler, then the result to a C-to-bytecode compiler, and the bytecode to an arithmetic interpreter built-in to the \texttt{fri} code. The interpreter has access to system-provided dynamic loading features, and through those it gains the capability to call up compiled programs, and individually access every subroutine and global variable defined in them. For the user this implies that an interpreted session can transparently \cod{USE} (\S\ref{USE}) or \cod{\#include} (\S\ref{C_interface}) compiled library modules, or even external C or FORTRAN functions, just as easily as a compiled program can.

\section{Program structure\label{Syntax}\index{syntax}}


The general structure of a \CPL\ program is composed of nested blocks of statements which take effect in
a hierarchical top-down order, much like in Pascal or Modula-2. Statements are
composed of keywords, identifiers and separators.

User-declared identifiers can contain alphabetic and numeric characters plus
\cod{\_} and can be of any length. Upper and lower case are considered distinct. The first character of an identifier must be non-numeric.

Separators can be spaces, tabs or comments (see \S\ref{Comments}) in any number and
sequence; at least one separator is required between contiguous words that
could otherwise be mistaken for a single identifier. Indenting is advised, to highlight statement blocks, but not required. The included text editor provides auto-indenting.

Statements within a block are delimited by either newline or \cod{;}, in any number and sequence with separators, which allows
for multi-statement lines without requiring (but permitting) a \cod{;} when the
statement ends at the end of a line. Multi-line statements can be broken after any keyword that necessitates of a further operand (\eg a \cod{,}): if the given line does not make sense as is, the compiler will automatically treat newline as a space and try to complete the statement on the next line. Line continuation can also be explicitly requested wherever a space is allowed by putting a \cod{\textbackslash} as the last character before newline.

Round, square and curly brackets can be used interchangeably wherever a
parenthesis is needed, in either expressions, array arguments or function arguments (see \S\ref{Parentheses}), but must be closed by a bracket of matching kind.

\cod{SUBROUTINE} and \cod{FUNCTION} declarations (see \S\ref{Functions}) are on the same
footing as \cod{VARIABLE}, \cod{CONSTANT} and \cod{TYPE}
declarations. Contrary to some other languages, declarations and executable statements  may be freely
intermixed in the program flow. The scope of a declaration extends from the
position where the declaration appears down to the end of the containing block.
It is recommended that declarations appear as close as possible to the position
where the declared item is used for the first time, as this enables the
compiler to catch a larger number of errors automatically. In most situations,
a variable can be declared and initialized in a single statement.

Loops and all other flow control statements (see \S\ref{Control}) implicitly define
new blocks within them, just as subroutine declarations do. A block can also be
explicitly delimited as a \cod{MODULE} (see \S\ref{MODULE}). A block can contain every declaration
that the outermost block can, including further subroutines\footnote{Nested subroutines and functions, however, appear as such in the generated C code, a feature that \texttt{gcc} can handle but is not in ANSI C.} and modules;
the declared items cease to exist on exit from the block.

\subsection{Comments\label{Comments}\index{Comments}}

Comments delimited by a \cod{!} without a following bracket extend to the end of line.

Comments delimited by \cod{!(}\ldots\cod{!)} or \cod{![}\ldots\cod{!]} or \cod{!\{}\ldots\cod{!\}} pairs can be inserted
wherever a space can, and can extend over a few characters or over several
lines. The compiler will match \cod{!\emph{bracket}} pairs within comments, and
consider the comment finished only when all opened such parentheses have been
closed.

\subsection{Parentheses\label{Parentheses}\index{Parentheses}}

Round, square and curly brackets can all be used interchangeably wherever a
parenthesis is needed, but the compiler requires that every opening bracket be
closed by a matching bracket of the same kind, just as in mathematics.
Alternating the three kinds of bracket, while not mandatory, makes for easier
reading by the programmer and more effective error catching by the compiler.
The included text editor automatically alternates brackets if requested.

When round brackets are used in statement prototypes in this manual, it is always understood that the other two styles may be used as
well. In such prototypes square brackets denote optional clauses, straight-up type fixed keywords, and slanted type variable fields or example identifiers that must not be written verbatim but suitably filled in.

\subsection{Macro definition and conditional compilation\label{C_preprocessor}\index{Macro definition and conditional compilation}}

C-preprocessor-like \cod{\#define} and \cod{\#if} constructs 
are available in a \CPL\ program to define macros and alter the flow of compilation. The \cod{\#include} directive is reserved for the inclusion of external C code (see \S\ref{C_interface}).
For the inclusion of \CPL\ files, the \cod{USE} (see \S\ref{USE}) and \cod{INCLUDE} (see \S\ref{INCLUDE}) statements are available.

\subsection{Source file\label{Source_file}\index{Source file}}

A \CPL\ program file directly consists of a sequential block of declarations and
statements, some of which may in turn contain other blocks, without an explicit
mark of the beginning and end of the complete program other than by the beginning and end of the file itself.

\subsection{USE\label{USE}\index{USE}}

Auxiliary source files can be used, and separately compiled if they were not already, through the statement
\begin{code}
USE \emph{filename}
\end{code}
The \texttt{.cpl} suffix is automatically appended if \cod{\emph{filename}} contains no dot. 
cod{\emph{filename}} can be relative or absolute as per unix convention. In addition,
the directory where the \cod{USE}d file resides becomes the Present Working Directory for the purpose of other file
references nested inside it.

The \cod{USE}d file is logically treated as if it were inserted in the program
sequence in place of the \cod{USE} statement, i.e. just as if \cod{INCLUDE} (see below) were
in its place. However, in reality the file is either compiled into a separate
object file or, if a more recent object file of the same name already exists,
not compiled at all. In the latter case the \CPL\ compiler reads the outermost
block only from the \cod{USE}d source file and skips the inner-block contents
altogether, generating the same effect as a Modula-2 DEFINITION MODULE or a C
header file.

Notice that the \cod{USE} statement does not define its own separate program block: declarations appearing at the outermost level in the \cod{USE}d file are global and stay in effect after the file is closed. The \cod{MODULE} (\S\ref{MODULE}) keyword is available to provide locality if desired.)

In case actual headers must be generated in order to save space (or to
distribute a library without source code) it is possible to \cod{USE} a source file
with empty subroutine bodies, since the compiler will not read the subroutine
bodies anyway, provided the object file carries a more recent
modification time (for instance, because it was compiled from a separate file, or just \emph{touch}ed --- by the unix \cod{touch} command --- afterwards.) If empty bodies are provided without adjusting timestamps, subroutines will be recompiled empty with obviously unwanted results.

In an interpreted program (to be processed by \texttt{icpl}) just as well, \cod{USE} loads a separately compiled module, whose subroutines can then be executed at compiled-code speed, and executes its main body, if any, immediately.

\subsubsection{File inclusion\label{INCLUDE}\index{File inclusion}}

Files included with

\begin{code}
INCLUDE \emph{filename}[(\emph{name}=\emph{value} [,\emph{name}=\emph{value}])]
\end{code}
are inserted in place of the \cod{INCLUDE} statement during
compilation, and are not separately compiled (as in \cod{USE}, see \S\ref{USE}). Nonetheless,
the directory of the \cod{INCLUDE}d file becomes the present working directory for
other references nested inside it.

If the optional arguments are present, \cod{\emph{name}} becomes an alias for \cod{\emph{value}} within the scope of the \cod{INCLUDE}d file.

\subsection{MODULE\label{MODULE}\index{MODULE}}

A block of code can be explicitly delimited as

\begin{code}
MODULE \emph{name}

\  \emph{some code}

END \emph{name}
\end{code}
or
\begin{code}
MODULE

\  \emph{some code}

END MODULE
\end{code}

This construction limits the visibility of variables declared inside the
\cod{MODULE} to within the module itself, and is the basis for information hiding. It replaces, and should be preferred to, subroutines that are called once only. An important difference is that variables declared in an outermost-scope \cod{MODULE} are static and retain their value across subroutine calls, even if they can only be accessed by statements and subroutines that exist inside the \cod{MODULE} itself. Notice also that source files recalled
with either \cod{USE} (see \S\ref{USE}) or \cod{INCLUDE} (see \S\ref{INCLUDE}) are not separately scoped unless
they are enclosed in a \cod{MODULE}. Control may be transferred to the end of a named
\cod{MODULE}, just as to the end of a \cod{SUBROUTINE}, by the \cod{EXIT \emph{name}} (see \S\ref{EXIT}) statement.

In order to be visible outside, a function declared inside a module must also
be declared before the \cod{MODULE} with the keyword \cod{FOLLOWS} (see \S\ref{FOLLOWS}). Externally visible variables or
constants can be simply declared before a \cod{MODULE} is entered, and assigned a value inside it when required.

\subsection{FUNCTION and SUBROUTINE\label{Functions}\label{FUNCTION}\label{SUBROUTINE}\index{Functions and Subroutines}}

Functions and subroutines have a long
and a short declaration form. In long form, a subroutine declaration reads

\begin{code}
SUBROUTINE \emph{name}(\emph{parameter declarations})

\  \emph{some code}

END \emph{name}
\end{code}

Correspondingly a function, i.e. a subroutine that returns a value and may
appear inside an expression, is declared as

\begin{code}
\emph{type} FUNCTION \emph{name}(\emph{parameter declarations})

\  \emph{some code}

END \emph{name}
\end{code}

In short form, a function whose body is composed of a single expression
may also be declared as

\begin{code}
\emph{type} FUNCTION \emph{name}(\emph{parameter declarations}) = \emph{expression}
\end{code}

A short syntax is also available for subroutines

\begin{code}
SUBROUTINE \emph{name}(\emph{parameter declarations}) = \emph{single-line block}
\end{code}
where \cod{\emph{single-line block}} is a sequence of statements separated by semicolons
and terminating at the first newline.

Parameters in the parameter list are by default separated by commas, but separators can also be customized (see \S\ref{Coactive_Parameter_Lists}). In a \cod{SUBROUTINE} call (but not in its declaration), brackets around the parameter list are optional and may be omitted.

The declared subroutine is visible from the point of its declaration to the
end of the enclosing block, just as all other declarations. This includes the
subroutine body itself, and recursive calls are thus automatically allowed.

A third alternative syntax, similar to the one used for  variables of FUNCTION type (see \S\ref{Subroutine_type}) has the result type denoted by the symbol "->" following the parameter declaration:

\begin{code} 
       FUNCTION \emph{name}( \emph{parameter declarations} )->\emph{type}

\  \emph{some code}

END \emph{name}
\end{code}
 
When this syntax is used, the result type may be, or contain, an ARRAY type
with dimensions depending on the formal parameters themselves (see \S\ref{array_parameters}). Any formal parameter as well may depend on the formal
parameters that precede it. For example:

\begin{code} 
       FUNCTION $massagearray$[ARRAY(*) OF REAL $v$]->ARRAY($v$.LO..$v$.HI) OF REAL
\end{code}

\subsubsection{Optional parameters\label{OPTIONAL}\index{Optional parameters}}

The parameter list of a function or subroutine can include optional items.      
In the function declaration, optional parameters must appear after all ordinary 
positional parameters (if there are any), separated by the keyword \cod{OPTIONAL}.
Each optional parameter must be assigned a default value in its declaration.    
Example:                                                                        

\begin{code}
SUBROUTINE test(INTEGER x; OPTIONAL INTEGER y=0,z=3; REAL w=3.14)              
\end{code}
           
Within the body of the function, optional parameters obey the same rules as local
constants or variables (if qualified \cod{VARIABLE}), just like ordinary      
parameters; on entry they acquire the value of the corresponding actual parameter if one is specified in the calling statement, or their default value otherwise.

In the calling statement, optional parameters follow all standard positional
parameters and can appear in any order (or not appear at all); they are
identified by name rather than by position. In the above example:

\begin{code}
test(3,w=1E2,y=1)
\end{code}

is a possible calling statement. Optional parameters cannot coact to disambiguate overloaded function names (see \S\ref{Coactive_Parameter_Lists}).

\subsubsection{Passing array parameters and their dimensions\label{array_parameters}\index{Array parameters}} 
Functions that accept array parameters of variable dimensions are a commodity
that is handled differently in different languages, especially when error
checking is desired. In CPL they are handled as follows.
 
The special dimension \cod{*} can be used to declare a POINTER TO ARRAY that
implicitly passes the bounds of the ARRAY together with its address (see \S\ref{ARRAY}). This is the most compact notation, as in \eg

\begin{code} 
       REAL FUNCTION NORM[ARRAY(*) OF REAL v]=SUM v(i)\^{}2 FOR ALL i
\end{code} 
 
 (same as the builtin \cod{NORM} function, see \S\ref{NORM}). When this notation is used, the
index bounds of the actual parameter are accessible in the function's body by
the \cod{LO} (see \S\ref{LO}) and \cod{HI} (see \S\ref{HI}) builtins (in the above example, these would be
\cod{v.LO} and \cod{v.HI}), and its number of elements by the \cod{LENGTH} (see \S\ref{LENGTH}) function, just
as for all \cod{ARRAY}s.

A formal parameter of type \cod{ARRAY(*)} also accepts a variable of its base type as its actual parameter, and transforms it into an ARRAY with dimension (0..0).

For more elaborate needs, for instance when two arrays must be enforced to have equal dimensions, \CPL\ allows previous formal parameters to be used in the declaration of
subsequent compound parameters. An example would be

\begin{code}
       SUBROUTINE $dosomething$[INTEGER $n$; ARRAY(1..$n$) OF REAL $a$,$b$,$c$]
       $\cdots$
       END $dosomething$
\end{code}

Within the body of such a subroutine or function, $a$, $b$ and $c$ are all assumed to
have the same dimensions 1..$n$; this is similar to the convention adopted in
languages (like C) that treat arrays like pointers with no special provision,
but with the difference that in \CPL\ the call of such a subroutine checks that
the actual parameters indeed have the specified dimensions at run time, and
throws an error if they don't and \cod{rtchecks} is on.
The same construction is available if
the dimension is not simply the $n$ parameter but an expression involving $n$, or
if the dimension is extracted from the bounds of a previous array as in

\begin{code}
       SUBROUTINE $dosomething$[ARRAY(*) OF REAL $a$; ARRAY($a$.LO..$a$.HI) OF REAL $b$,$c$
       $\cdots$
       END $dosomething$
\end{code}

Additionally, the return type of a \cod{FUNCTION} can as well be an \cod{ARRAY} involving a
formal parameter in its dimensions. An example is:
 
\begin{code}
       FUNCTION $dosomething$[ARRAY(*) OF REAL $a$,$b$]->ARRAY($a$.LO..$b$.HI) OF REAL
       $\cdots$
       END dosomething
\end{code}
 
After this declaration the dimensions of the return type are known to the
compiler, and can either be checked by \cod{rtchecks} or used to allocate a suitable \cod{ARRAY} \cod{CONSTANT}.

\subsubsection{Nested functions\label{Nested_functions}\index{Nested functions}}

Functions and subroutines exist on a par with variable declarations (see \S\ref{Declarations}), and
can appear anywhere in a block, another function included. Nested functions are
therefore an integral part of \CPL. Like all other block contexts, they have
access to variables (and functions) declared before them, in the same or in
an enclosing block. The present implementation of nested functions relies
on nested functions in the generated C code, and is therefore only compatible
with C compilers that handle nested functions as a nonstandard feature (\texttt{gcc} is
one of them).

\subsubsection{Function prototypes\label{Function_prototypes}\label{FOLLOWS}\index{Function prototypes}}

In case a subroutine or function must be called before its body occurs,  the construction

\begin{code}
SUBROUTINE \emph{name}(\emph{parameter declarations}) FOLLOWS
\end{code}
(or its equivalent for functions) can be used to pre-declare its calling syntax. This prototype declaration is always available, but only really needed for mutually referencing subroutines (as in Pascal or C) or for subroutines whose body is nested in a subsequent \cod{MODULE} (see \S\ref{MODULE}).

\subsubsection{EXIT\label{S_EXIT}\index{SUBROUTINE EXIT}}
Both \cod{SUBROUTINE} and \cod{FUNCTION} return to the calling procedure either at the \cod{END \emph{name}} (which also concludes the subroutine's body) or
\cod{EXIT \emph{name}} statement (which may appear in its interior).

\subsubsection{RESULT\label{RESULT}\index{RESULT}}
Within a function's body, a variable named
\cod{RESULT}
is implicitly declared with the type of the function's result, and can be used
just like any other local variable of that type within the body of the function;
the function returns the value of the \cod{RESULT} variable at the time of
exit. All primitive and compound types are allowed for the result of a function.
The statement

\begin{code}
RETURN \emph{value}
\end{code}
is a shorthand for

\begin{code}
RESULT=\emph{value}; EXIT \emph{name}.
\end{code}

The \cod{SUBROUTINE} and \cod{FUNCTION} keywords are also used to declare variables that may be used to call such procedures indirectly (see \S\ref{Subroutine_type}).

\subsubsection{INLINE functions and subroutines\label{INLINE}\index{Inline functions and subroutines}}

The keyword \cod{INLINE} prefixed to a \cod{SUBROUTINE} or \cod{FUNCTION}
declaration (in either long or short form) generates a compile-time macro that
is inserted in place of each call to the subroutine.

\subsection{C interface\label{C_interface}\index{C interface}}

The directive

\begin{code}
\#include "\emph{file}"
\end{code}
or
\begin{code}
\#include <\emph{file}>
\end{code}
in addition to being copied verbatim into the generated C code, activates a \CPL\ interface mechanism.
\cod{\emph{file}} is expected to contain a C header or program, which is scanned for
declarations by the \CPL\ compiler. The declarations and prototypes contained in
the C file thus become transparently available to (and enforced by) the including \CPL\ program.

An include directory path (similar to the \cod{-I} C compiler directive) can be
specified with

\begin{code}
\#includedir "\emph{path}"
\end{code}

The name of the corresponding object file or other linking options can be
transmitted to the linker with

\begin{code}
\#link "\emph{linking options}"
\end{code}

The following standard libc headers
\begin{itemize}\itemsep-0.3em
\item stdlib.h
\item stdio.h
\item fcntl.h
\item math.h
\item limits.h
\item string.h
\item setjmp.h
\item errno.h
\item signal.h
\end{itemize}
are pre-included in all \CPL\ programs and need not (but may) be included again.

\subsubsection{C SECTION\label{C_SECTION}\index{C SECTION}}

A C code excerpt may also appear directly in the middle of the \CPL\ source file if
encapsulated in a \cod{C SECTION} as follows:

\begin{code}
C SECTION

\  \emph{some C code}

END C SECTION
\end{code}

This has the same effect as recalling with \cod{\#include} a file that contains
\cod{\emph{some C code}}. That is, the identifiers and function prototypes declared inside the \cod{C SECTION} become transparently available to the \CPL\ program. \cod{\emph{some C code}} ends up in one and the same file with the C translation of the enclosing \CPL\ code.

As an alternative, C code may also be inserted between \cod{<*} and \cod{*>} delimiters, as in

\begin{code}
<* \emph{some C code} *>
\end{code}
In this usage, anything that appears between \cod{<*} and \cod{*>} is copied verbatim in
the generated C file without any intervention or symbol extraction by the \CPL\ compiler. It is up to
the programmer to ensure that the resulting code makes sense. In fact, this
construction was mostly introduced as a hack to be used when \CPL\ was still incomplete; its use is now discouraged.

\subsection{FORTRAN interface\label{FORTRAN_interface}\index{FORTRAN interface}}

An interface to separately compiled FORTRAN subroutines is provided by the statements

\begin{code}
FORTRANCALL \emph{subrname}(\emph{parameters})
\end{code}
and
\begin{code}
\emph{type} FORTRANFUNCTION \emph{functionname}(\emph{parameters})
\end{code}

These have the following effects:
\cod{\emph{subrname}} or \cod{\emph{functionname}} is copied into the generated C source with an appended underscore;
all arguments are passed by reference, according to FORTRAN convention;
any arrays are replaced by a reference to their first element.

The user must take care of the fact that FORTRAN allocates memory for multidimensional arrays with
indices in the reverse order with respect to both C and \CPL. Arrays to be
passed as such arguments should be declared and used accordingly.

For (e.g. lapack) library calls that require the number of elements used in
actual memory storage for the rows of a matrix as an argument, the
pseudofunction

\begin{code}
STRIDEOF(\emph{array})
\end{code}
provides the argument to be passed.

Please be aware that argument type checking is suppressed during a \cod{FORTRANCALL} (as the
necessary information is not available). Type checking can be restored,
however, by wrapping every \cod{FORTRANCALL} in a suitable \cod{INLINE} \cod{SUBROUTINE} (see \S\ref{INLINE}).

The name of the object file providing the FORTRAN compiled subroutines must be
transmitted to the linker, just as in \S\ref{C_interface}, through the directive
\begin{code}
\#link "linking options"
\end{code}
together with any other linking options required (\eg \texttt{-lgfortran} to provide the FORTRAN run time library, if \texttt{g77} was used as the FORTRAN compiler).

\section{Variables, constants and types\label{Declarations}\index{Variables, constants and types}}

Variables, constants and types are denoted by identifiers introduced through
the relevant declarations as follows. Identifiers cannot be duplicated as long
as a previous declaration is in effect (in either the same or an enclosing
block). An identifier which was not previously declared in the same or an
enclosing block and does not coincide with a \CPL\ keyword (as listed in the \hyperlink{Keyword_index}{Keyword index}) is denoted as \cod{\emph{newid}} below.

A general declaration for a variable or a constant has the form:

\begin{code}
[\emph{constorvar}] \emph{typedeclarator} \emph{vardeclaration} [, \emph{vardeclaration}]
\end{code}
where \cod{\emph{vardeclaration}} denotes

\begin{code}
\emph{newid}[\emph{postfix modifier}][= \emph{value}]
\end{code}
and \cod{\emph{constorvar}} may be either \cod{CONSTANT} or \cod{VARIABLE} or blank.

\CPL\ is a type-checking language and requires that every \cod{VARIABLE} be declared
before it is used. It can be initialized at the same time by the optional
\cod{= \emph{value}} field.

\subsection{VARIABLE\label{VARIABLE}\index{VARIABLE}}

The keyword \cod{VARIABLE}, put either before or after the \cod{\emph{typedeclarator}} in a declaration, specifies that the declared item is a
regular variable and can be modified multiple times. This is, in fact, the default and the word
\cod{VARIABLE} is optional in ordinary declarations.

However, subroutine formal parameters are by default \cod{CONSTANT}, with the
actual parameters as values, and must be explicitly declared \cod{VARIABLE} if they are to be reassigned within the subroutine's body (thus becoming different from the value of their actual parameter which stays unchanged).

Notice that the value assigned to a \cod{VARIABLE} parameter is lost after the
subroutine returns. In order to change the value of variables which exist
outside the subroutine, the corresponding formal parameter must be declared
 a \cod{POINTER}. Indeed \cod{VARIABLE} behaviour of a formal parameter is seldom
needed, and making \cod{CONSTANT} the default improves error catching.

When a \cod{VARIABLE} identifier is used inside an expression, it initially stands for
the variable's address in memory, and thus can be used wherever a \cod{POINTER} is
expected; the same identifier gets automatically converted to the variable's
value where needed through Concealed Pointer Lookup (see \S\ref{Concealed_Pointer_Lookup}).

\subsection{CONSTANT\label{CONSTANT}\index{CONSTANT}}

The keyword \cod{CONSTANT}, put either before or after the \cod{\emph{typedeclarator}} in a declaration, specifies that the declared item (usually assigned a value in the same statement)
cannot be modified later. Thus a \cod{CONSTANT} is not the same thing as a ``compile-time
constant'', that is an expression whose value can be known at compile time, but
more generally a value which is prohibited from appearing on the left side of an assignment or having its address assigned to a \cod{POINTER}.

The value of the \cod{CONSTANT} is usually specified in the declaration itself but
occasionally this specification may be deferred. When this happens the constant declaration takes the role
of a constant prototype, and may be optionally followed by the keyword \cod{FOLLOWS}. (For instance, a constant may be declared outside a \cod{MODULE},
but its value only be known inside the module itself; or the value may be \cod{READ} as input.) If the \cod{CONSTANT} is not initialized in the declaration itself, it is enforced to only be assigned once in the course of its block.

As a shortcut, a statement of the form
\begin{code}
\emph{newid} = \emph{value}
\end{code}
implicitly declares a \cod{CONSTANT} whose type and value are the type and value of the \rhs\  (see, however, \S\ref{Concealed_Pointer_Lookup} if the
address of the r.h.s. is desired instead).

If the \rhs\ of a \cod{CONSTANT} declaration is known at compile time (what is
denoted as a compile-time constant), the new identifier becomes just an alias
for it and no storage is assigned. 

An alias for any expression, even one which does not represent a compile-time
constant, can always be defined through the statement

\begin{code}
\emph{newid} == \emph{expression}
\end{code}

See \S\ref{Deferred_assignment}.

\subsection{TYPE\label{TYPE}\index{TYPE}}

A new type identifier can be declared as

\begin{code}
[TYPE] \emph{newid} = \emph{typedeclarator}
\end{code}
and be used in subsequent declarations in the place of the
\rhs\ The \cod{TYPE} keyword is optional, since the presence of a type declarator on the \rhs\ already distinguishes this declaration from an assignment.

\subsubsection{TYPEOF\label{TYPEOF}\index{TYPEOF}}

The type of an already declared variable or expression can be recovered through the pseudo-function

\begin{code}
TYPEOF(\emph{variable})
\end{code}
which can appear wherever a type declarator can (\eg it can be assigned to a new type
identifier or directly used to declare further variables, or in the construction
of a compound type). When applied to a \cod{POINTER} or \cod{STORED} (see \S\ref{STORED}) type,
this function returns the underlying base type.

In order to test whether a variable is of a given type the \cod{IS}
comparison operator is available (see \S\ref{IS}).

\subsubsection{SIZEOF\label{SIZEOF}\index{SIZEOF}}
The memory size in bytes of a type or variable (useful when using the C interface of \S\ref{C_interface} or also to estimate the memory requirements of the program) can be
obtained from the pseudo-function

\begin{code}
SIZEOF(\emph{type or variable})
\end{code}

\subsubsection{Postfix type modifier\label{Postfix}\index{Postfix type modifier}}

A postfix \cod{\^{}} denotes either pointer dereferencing (see \S\ref{Concealed_Pointer_Lookup}) or a short form of the \cod{POINTER} declaration. For instance

\begin{code}
INTEGER a\^{},b
\end{code}
is a shorthand for
\begin{code}
POINTER TO INTEGER a; INTEGER b
\end{code}

A postfix form of the \cod{ARRAY} declaration
\begin{code}
ARRAY(1..10) OF REAL a,b,c
\end{code}
is
\begin{code}
REAL a(1..10),b(1..10),c(1..10)
\end{code}

When both a postfix modifier and a prefix type declarator are simultaneously
present in a declaration, they represent the compound type that is obtained by
prepending the effect of the postfix modifier to the explicit type declarator. For example

\begin{code}
STRUCTURE(INTEGER x,y) A\^{}(10)
\end{code}
is equivalent to
\begin{code}
POINTER TO ARRAY(10) OF STRUCTURE(INTEGER x,y) A
\end{code}

\subsubsection{Type identity\label{Type_identity}\index{Type identity}}

When are two \cod{VARIABLE}s of the same \cod{TYPE}? Pascal-like strongly typed languages
only consider variables of the same type when they are declared by the same
typename or they appear in one and the same compound declaration. In other words,
repeated compound declarations are not considered to be of the same type even when they
are equal, just as two different typenames
with similar definitions are not considered equal. C, on the other hand, handles typenames as bare
aliases, and compares their definitions directly. None of these languages has
runtime-dimensioned arrays. \CPL\ takes the C approach when possible, and
equates a given  typename to its definition, but considers two explicitly given typenames
different even when they have equal definitions. Generally, it is
good practice to declare variables that must be of the same type in a single
declaration, or through a single typename, or through the \cod{TYPEOF} pseudo-function (see \S\ref{TYPEOF}). This is
particularly true with runtime-dimensioned arrays (\cod{ARRAY}s whose bounds are
specified by a \cod{VARIABLE} and only known at run time). In fact, since a
\cod{VARIABLE} may take on different values in different parts of the program,
multiple declarations of this type are necessarily handled as being different.

In order to test whether a variable is of a given type the \cod{IS} comparison operator is available (see \S\ref{IS}).

\subsection{Implicit declarations\label{Implicit_declaration}}

The declaration of a \cod{CONSTANT} or \cod{TYPE} may be
implicit:

\begin{code}
\emph{newid} = \emph{value}
\end{code}
declares a new \cod{CONSTANT}, whose value cannot be modified
later in the program and whose type is the type of the \rhs

\begin{code}
\emph{newid} = \emph{typedeclarator}
\end{code}
declares a new \cod{TYPE} identifier to denote the
(possibly compound) type on the \rhs

A \cod{\emph{newid}} used as the running index of a \cod{FOR} (see \S\ref{FOR}) loop is implicitly declared a \cod{CONSTANT} for the scope of the loop block.

\subsubsection{FOLLOWS}
A declaration of the form
\begin{code}
CONSTANT \emph{newid} FOLLOWS
\end{code}
defines a  \cod{CONSTANT} whose value is hidden (typically, inside a subsequent \cod{MODULE}) or to be read (once only) from a file.

\subsection{Primitive types\label{TYPEs}\index{Primitive types}}
Types can be primitive or compound. Compound types are specified through a \emph{compound-type expression}, assigned to a type identifier or directly used as the type declarator (common to all variables) in a
declaration list, or through a \emph{postfix modifier} attached to each individual
variable to which it is meant to apply.

Primitive types are:

\subsubsection{BOOLEAN\label{BOOLEAN}\index{BOOLEAN}}

Primitive type declarator for boolean logical variables, translated into C type
\texttt{int}.

\cod{BOOLEAN} values are denoted as \cod{YES} or \cod{TRUE} and \cod{NO} or \cod{FALSE}, and written as ``Y''
and ``N'' by \cod{WRITE} instructions. On \cod{READ}ing, any complete word starting with ``T'', ``Y'', ``t'' or ``y'' is
read as \cod{YES}, any word starting with ``F'', ``N'', ``f'', ``n'' as \cod{NO}.

See also \S\ref{Comparison}, \S\ref{BOOLEAN_ops}, \S\ref{BOOLEAN:builtin}.

\subsubsection{CHAR\label{CHAR}\index{CHAR}}

Primitive type declarator for character variables, translated into C type
\cod{char}.

Where needed, a single-character \cod{STRING} literal is implicitly converted
to a \cod{CHAR} constant, which is in turn implicitly converted to its \cod{INTEGER} ascii
code. \cod{INTEGER} to \cod{CHAR} conversion must be explicit (see \S\ref{type_conversion_builtin}).

\subsubsection{INTEGER\label{INTEGER}\index{INTEGER}}

Primitive type declarator for integer variables, translated into C type \cod{int}.

Can be implicitly converted to a \cod{REAL}. \cod{REAL} to \cod{INTEGER} conversion must be
explicit (see \S\ref{type_conversion_builtin}).

\subsubsection{REAL\label{REAL}\index{REAL}}

Primitive type declarator for real variables, translated into C type \cod{double}.

\subsubsection{SINGLE\label{SINGLE}\index{SINGLE}}

Primitive type declarator for single-precision real variables, translated into
C type \cod{float}.

Can be implicitly converted to a \cod{REAL}. \cod{REAL} to \cod{SINGLE} conversion must be explicit
(see \S\ref{type_conversion_builtin}).

\subsection{Enumerated type ENUM\label{ENUM}\index{ENUM}}
Not really a compound type, \cod{ENUM} variables (much like C enum or Modula2 enumerations) take a limited
set of explicitly enumerated values. The syntax of this type declaration is

\begin{code}
       ENUM($id1$,$id2$,\ldots)
\end{code}

and can directly declare variables or be assigned to a new type name like all
other \cod{TYPE}s. $id1$, $id2$, etc. are thereby declared as constants to which
variables of the given type can be set or compared. The purpose of \cod{ENUM}
variables is to flag a limited set of alternatives, while forbidding
out-of-range values which could arise if \cod{INTEGER}s were used for this purpose.
No other operations than assignment and comparison for equal are defined on \cod{ENUM}s.

\subsection{STRUCTURE\label{STRUCTURE}\index{STRUCTURE}}

The type declarator for structures is

\begin{code}
STRUCTURE(\emph{field declarations})
\end{code}

Field declarations follow exactly the same syntax as general variable
declarations. For instance, to declare a structure containing one integer field
named \cod{i} and two real fields named \cod{x} and \cod{y}, one would write:

\begin{code}
STRUCTURE(INTEGER i; REAL x,y)
\end{code}

Structure elements are equally selected by either the traditional dot notation

\begin{code}
\emph{structure}.\emph{field}
\end{code}
or the, less usual but equivalent, selector-function notation

\begin{code}
\emph{field}(\emph{structure})
\end{code}

Symmetrically any function of one or more arguments can be called either as

\begin{code}
\emph{function}(\emph{firstargument}[,\emph{morearguments}])
\end{code}
or as
\begin{code}
\emph{firstargument}.\emph{function}[(\emph{morearguments})]
\end{code}
(a notation apt to appeal to those who are accustomed to object-oriented languages).

A structure declaration may also contain one or more anonymous fields. If a
type declarator appears alone, without any following variable declaration, an
anonymous field is declared, to which the given structure can be implicitly
converted. For instance, after the declaration:

\begin{code}
TYPE t1=STRUCTURE(REAL a,b)

STRUCTURE(t1; INTEGER n) var
\end{code}
the structure \cod{var} includes a field of type \cod{t1} of which \cod{var.a} denotes the \cod{a} field.

Anonymous structure fields are \CPL's way of defining objects that inherit the
structure of other more general objects and add their own specific fields. When
these objects are accessed as arguments to functions or subroutines, the
appropriate function among those with the same name and different arguments is
automatically selected. See also \S\ref{Coactive_Parameter_Lists} and \S\ref{DYNAMIC}.

\subsubsection{Implicit access to structure fields and functions\label{WITH}\index{Implicit access to structure fields and functions}}

Like in Pascal or Modula2, a field of a structure may be implicitly selected by the field's name alone when the structure's name has been specified in a \cod{WITH} statement. \cod{WITH} has both a multi-line form:

\begin{code}
WITH \emph{some structure} [,\emph{some other structure}] [:]

\  \emph{code block}

[END WITH]
\end{code}
and a single-line form:

\begin{code}
WITH \emph{some structure} [,\emph{some other structure}] : \emph{single-line block}
\end{code}

The multi-line form is recognized from a newline following the
structure list or the optional colon; in this case the scope of the statement
extends either to the optional \cod{END WITH} statement or to the end of the
enclosing block; in the second to the end of line. Within the scope of a \cod{WITH}
statement the given structure (or structures) is implied wherever a field
name alone appears. For instance, after the declaration

\begin{code}
STRUCTURE(REAL x,y) v1,v2
\end{code}
writing
\begin{code}
WITH v1: x=v2.y; y=3
\end{code}
becomes equivalent to \cod{v1.x=v2.y; v1.y=3}, or to \cod{x(v1)=y(v2); y(v1)=3}.

Standing the symmetry of treatment of structure fields and functions by \CPL\ (see \S\ref{STRUCTURE}), a function argument may also be specified in a \cod{WITH} clause. Thus, for instance,

\begin{code}
WITH 3: WRITE SIN
\end{code}
is equivalent to \cod{WRITE SIN(3)}. Of course this notation is not very useful for \cod{REAL} functions, but it becomes handy when functions are defined for the purpose of accessing user-defined compound types as though they were structures.

\subsubsection{Variable size STRUCTURE\label{variable_size}\index{variable size STRUCTURE}}
\cod{STRUCTURE}s may be defined that contain \cod{ARRAY}s the size of which will only be
known at run time. Such \cod{STRUCTURE} \cod{TYPE}s are declared with one or more \cod{INTEGER}
\cod{CONSTANT}s among their fields. When a variable of the given type is allocated,
either by using the name assigned to its \cod{TYPE} in a declaration (see \S\ref{Declarations}) or in a
\cod{NEW} (see \S\ref{NEW}) statement, the values following the type name in parentheses (like
arguments of a function) are assigned to these \cod{CONSTANT}s by order, and are
simultaneously used to allocate a chunk of memory of the appropriate size.
Being \cod{CONSTANT}s, these fields cannot be altered in the subsequent life of the
allocated \cod{STRUCTURE}. Example:

\begin{code}
      TYPE $twoarrays$=STRUCTURE[REAL $x$; INTEGER CONSTANT $n$
              POINTER TO $twoarrays$ $next$
              ARRAY(1..$n$) OF INTEGER $k$; ARRAY(-$n$..2*$n$) OF REAL $data$]
      $\cdots$
      $twoarrays$(8) $mytwoarrays$
      $mytwoarrays$.$next$=NEW $twoarrays$(2*$mytwoarrays$.$n$)
\end{code}

Notice that this syntax is intentionally similar to the syntax for $FUNCTION$s
that contain $ARRAY$s of runtime-specified size as arguments. Compare \S\ref{array_parameters}. In fact, the set of parameters of a given function may
usefully be thought of as a structure, although one that is invisible to the
programmer and only exists as a conceptual model.

\subsection{ARRAY\label{ARRAY}\index{ARRAY}}

The prefix type declarator for arrays is

\begin{code}
ARRAY(\emph{dimension}[,\emph{dimension}]) OF \emph{type}
\end{code}
The postfix declarator is just \cod{(\emph{dimension}[,\emph{dimension}])}. Each dimension is
specified either as \cod{\emph{lower bound}..\emph{upper bound}} or as \cod{\emph{upper bound}} only, in
which case the lower bound is taken to be unity. Specifying multiple dimensions,
like in
\begin{code}
{ARRAY(\emph{dimension1},\emph{dimension2}) OF \emph{type}}
\end{code}
is a shorthand for
\begin{code}
ARRAY(dimension1) OF ARRAY(dimension2) OF \emph{type}
\end{code}

Therefore the order in which the elements of the array are stored in memory is
similar to C and Pascal and different from FORTRAN.

Variable dimensions are allowed wherever (compile-time) constant dimensions are; the compiler
automatically decides for static or dynamic allocation of memory space as
needed. Arrays can contain elements of any primitive or compound type and can
in turn appear as fields in structures or be pointed to by pointers.

The special dimension \cod{*} can only appear in the declaration of an \cod{ARRAY} formal
parameter (see \S\ref{array_parameters})  or of a \cod{POINTER TO ARRAY}, and declares a special type of pointer that stores the current dimensions of the array together with its address. A
formal parameter or pointer of this type can be dynamically assigned arrays
of different length and still provide runtime range checking and \cod{FOR ALL} (see \S\ref{ALL}) looping. The current upper and lower bounds of an array
can always be recovered through the \cod{HI} and \cod{LO} functions,
and the total number of elements through the \cod{LENGTH} function (see below). In the
case of multidimensional arrays, these functions apply to the first index only;
subarray selection (see \S\ref{subarray}) can be used in order to extract the bounds of other
indices. As a shortcut, \cod{LO1}, \cod{LO2}, \cod{LO3} and \cod{HI1}, \cod{HI2}, \cod{HI3} can be used to refer to
the first three indices.

\cod{ARRAY} elements are accessed through \cod{INTEGER} indices in brackets as usual, \eg

\begin{code}
REAL arr(1..8,2..9)

WRITE arr(4,5)
\end{code}

However, all three kinds of bracket are allowed, as everywhere in \CPL\ (see \S\ref{Parentheses}). It should be noted that the above declaration is a shorthand for

\begin{code}
ARRAY(1..8) OF ARRAY(2..9) OF REAL arr
\end{code}
Therefore, a notation such as \cod{arr(4)} is allowed, and has type \cod{ARRAY(2..9)}\cod{OF
REAL}. This is a particular case of a subarray (see \S\ref{subarray}). By the same token, the
second line might also have been written as

\begin{code}
WRITE arr(4)(5)
\end{code}

The following special functions are defined on \cod{ARRAY}s:

\subsubsection{HI\label{HI}\index{HI}}

The function \cod{HI} applied to an \cod{ARRAY} argument returns the upper bound of its first index.
\cod{HI1} is a synonym for \cod{HI}; \cod{HI2} and \cod{HI3} return the upper bound of the second and
third index respectively. Further indices can be accessed by subarray selection (see \S\ref{subarray}).

The argument of \cod{HI} is implicit (\ie \cod{HI} may be written without any argument as if a \cod{WITH} --- see \S\ref{WITH} --- were in action) when \cod{HI} is used inside the index itself as in
\cod{arr(HI-3)} or in the specification of a \cod{FOR} (see \S\ref{FOR}) loop.

\subsubsection{LO\label{LO}\index{LO}}

The function \cod{LO} applied to an \cod{ARRAY} argument returns the lower bound of its first index.
\cod{LO1} is a synonym for \cod{LO}; \cod{LO2} and \cod{LO3} return the lower bound of the second and
third index respectively. Further indices can be accessed by subarray selection (see \S\ref{subarray}).

The argument of \cod{LO} is implicit (\ie \cod{LO} may be written without any argument as
if a \cod{WITH} --- see \S\ref{WITH} --- were in action) when \cod{LO} is used inside the index itself as in
\cod{arr(LO+2)} or in the specification of a \cod{FOR} (see \S\ref{FOR}) loop.

\subsubsection{LENGTH\label{LENGTH}\index{LENGTH}}

The function \cod{LENGTH} applied to an \cod{ARRAY} argument returns its number of elements. \cod{LENGTH}
is equivalent to \cod{HI-LO+1}. In a multidimensional \cod{ARRAY}, the range of
the first index only is returned.

\subsubsection{STRING\label{STRING}\index{STRING}}

Type declaration \cod{STRING} is a synonym for \cod{ARRAY(*) OF CHAR}, and can be
used wherever the last can. In particular, a \cod{STRING} can be subscripted or 
a subarray extracted from it, and its bounds and length are available through functions \cod{LO}, \cod{HI}, and \cod{LENGTH}. \cod{STRING} valued constants (which are always indexed starting from 0) can be generated either at compile time, as quoted strings of characters (see below), or dynamically through string concatenation (see \S\ref{String}), and can either be given a name through an implicit declaration (see \S\ref{Implicit_declaration}) or passed as actual parameters to function calls and further string concatenations.

Quoted strings of characters, also known as literals, are delimited by either
double \cod{"} or single \cod{'} quotes, in matching pairs. A single quote may appear
inside a literal delimited by double quotes and vice versa. In addition, C
escape sequences are literals, but contrary to C must be unquoted.  The empty
string \cod{""} is also a valid (zero-character-long) \cod{STRING} value.

Finally, in unix-shell-like syntax, multi-line literals may be introduced by \cod{<{}<\emph{delimiter}} and
ended by \cod{\emph{delimiter}} alone on a new line, where \cod{\emph{delimiter}} can be any character sequence that is
not to be part of the literal itself. This is useful in programs that output multiline constant strings, for instance web servers. Within such multiline strings, the sequence \cod{\emph{delimiter variable delimiter}} (using the same \emph{delimiter} defined for the whole string, but without any intervening \texttt{newline}) represents a variable that will be treated as though it appeared in a \cod{WRITE} statement (see also \S\ref{String}).

Where appropriate, a single-character literal is implicitly converted to a \cod{CHAR}
constant, which is in turn implicitly converted to an \cod{INTEGER} ascii code.
Strings may also be concatenated among themselves and with variables (see \S\ref{String}).

Standard C functions returning a \cod{char*} result are also interpreted as generating a \cod{STRING} value. Just
as for all C interface (see \S\ref{C_interface}) calls, it is the programmer's responsibility to free any possibly \cod{malloc}-ed space where appropriate for such functions.

As an experimental feature, and an exception to the general rule for \cod{ARRAY(*)}, \cod{STRING} variables may also be declared and assigned a value.

\subsection{Subarray selection\label{subarray}\index{Subarray selection}}

A portion of an array can be selected by specifying new dimensions. For
example, if a variable \cod{arr} is declared as \cod{REAL arr(1..10)}, the notation

\begin{code}
arr(2..5)
\end{code}
denotes an array formed by the elements from 2 to 5 of array \cod{arr} (which must lie in the range of declared dimensions). The whole range can also be selected, by specifying \cod{*} as the subrange. As a more involved example, if the declaration is
\begin{code}
REAL arr(1..7,-2..2,3..8)
\end{code}
then
\begin{code}
arr(2..5,*,6)
\end{code}
denotes a two-dimensional array formed by selecting the range 2..5 of the first
index and the whole range of the second index of array \cod{arr}, with the third
index set equal to 6.

The most general form of subarray selection can also specify an offset and one or more strides. The notation for this is:

\begin{code}
arr(a+b1*(\emph{newdim})[+b2*(\emph{newdim})...])
\end{code}

where \cod{a}, \cod{b1}, \cod{b2}\ldots are \cod{INTEGER}s and \cod{\emph{newdim}} is either a subrange like \cod{\emph{l}..\emph{h}}, or \cod{*}.
The meaning of this notation is that the subarray must be formed of those
elements of the original array \cod{arr} whose index is the specified linear function
of the new index or indices. The latter are either constrained in the
explicitly indicated range \cod{l..h} or in the maximum admissible range \cod{*}. In fact
a general linear expression may be specified, with terms in any order and the
possible occurrence of parentheses.

Another form of subarray selection is index permutation, which in the
particular case of a two-dimensional array becomes matrix transposition.

\begin{code}
arr(*\emph{n},...)
\end{code}

denotes the array obtained from \cod{arr} by making the \emph{n}$^{th}$ index become the first
and shifting forward all the others. As a particular case, the transpose of a two-dimensional matrix \cod{A} is
\cod{A(*2,*)}, or just \cod{A(*2)}; the alternative notation
\begin{code}
TRANSPOSED(A)
\end{code}
is also provided.

Finally, a subarray can be extracted from an \cod{ARRAY} of \cod{STRUCTURE}s. If the
declaration is, for instance,

\begin{code}
ARRAY(1..10) OF STRUCTURE(INTEGER i; REAL r) arr
\end{code}
then by a straightforward notation \cod{arr.r} is the \cod{ARRAY(1..10) OF REAL} composed
of all the \cod{r} fields of each element of \cod{arr}. Thus, for instance, \cod{arr(5).r} and
\cod{arr.r(5)} denote the same \cod{REAL} field. \cod{arr.r} is an \cod{ARRAY} in its own right; it can
be copied, passed as a parameter or subjected to further subarray selections.

It should be noted that all subarray selections are just address manipulations (in fact, the most general \emph{linear} address manipulations):
no element of the array is ever moved and the selected subarray can appear on
the left as well as on the right side of an assignment and can be passed as a
\cod{POINTER TO ARRAY(*)} parameter to a function or subroutine. No computational
penalty is incurred because of the sheer size of either the original or the
selected array, but on the other hand any modifications of the subarray also
affect the original.

All subarray selections can also be applied to \cod{STORED ARRAY}s (see \S\ref{STORED}) and \cod{STRUCTURED ARRAY}s (see \S\ref{STRUCTURED}.

\subsection{Compound index\label{compound_index}\index{Compound index}}

\CPL\ allows a multidimensional \cod{ARRAY} to be accessed
sequentially through a single index, for instance in order to apply linear
algebra to it. The syntax for this operation is a sequence of as many \cod{*}s as indices are to be compounded. For example if \cod{V} is defined as

\begin{code}
ARRAY(0..4,1..5,-7..7,0..2) OF REAL V
\end{code}
then \cod{V(****)} is a onedimensional \emph{view} of the same array (not to be confused with \cod{V(*,*,*,*)} which represents its original arrangement). By the same token
\cod{V(*,**,*)} is the three-dimensional array obtained by compounding the two
central indices into one. Bounds of a compound index can be retrieved by the
usual \cod{LO} (see \S\ref{LO}) and \cod{HI} (see \S\ref{HI}) functions.

Compound indices are restricted to arrays (or portions of them) that are
contiguous in memory. This sometimes precludes their use after subarray
operations that permute, stride, or restrict the range of indices. When this is the case, a compile-time error results.

\subsection{Linear-space ARRAY operations\label{ARRAY_ops}\index{Linear-space ARRAY operations}}

Linear-space operations are supported on congruent \cod{ARRAY}s of
any number of dimensions and a base type on which arithmetic operations are defined. Such arrays can be added to or subtracted from each other and
multiplied or divided by a scalar inside an expression, as well as appear on
the left side of an assignment.

The scalar product of two congruent arrays is denoted by the infix operator \cod{|}.
This is defined as the sum over all dimensions of the products of corresponding
elements. (For \cod{COMPLEX} the first operand is conjugated first.)

The constant \cod{0} (but not any other constant) can be assigned as a value to a whole array.

Some builtin (see \S\ref{Builtin}) functions operate on whole arrays. A larger set of vector and matrix algebra operations are supported by the \texttt{rbmat.cpl} and \texttt{cbmat.cpl} libraries.

Matrix transposition was discussed in \S\ref{subarray}.

All \cod{ARRAY} operations can also be applied to \cod{STRUCTURED ARRAY}s (see \S\ref{STRUCTURED}) and \cod{STORED ARRAY}s (see \S\ref{STORED}).

\subsection{Implicit looping over array indices\label{Einstein_convention}\index{Implicit looping over array indices}}

When indices of arrays appearing in an expression are prefixed by a \cod{\$} sign, the Einstein convention is
implied\footnote{the \texteuro\ (euro) symbol might have been appropriate to match the ``E'' of
Einstein, but is not part of the 7-bit ascii set. Thus we exchanged it for American currency.}
. That is, in a product (a contraction) the Einstein index must appear
exactly twice, and the result will be summed over all values of this
index; in a sum and assignment, Einstein indices must appear exactly once in
each and all terms including the \lhs, and the assignment will be repeated
over all values of such indices. For example:

\begin{code}
C(\$i,\$j)=A(\$i,\$j,\$m)*B(\$i,\$m,\$j)+D(\$j,\$i)
\end{code}

\subsection{STRUCTURED ARRAY\label{STRUCTURED}\index{STRUCTURED ARRAY}}

A \cod{STRUCTURE} (see \S\ref{STRUCTURE}) of like elements that may also be accessed as a
one-dimensional \cod{ARRAY} (see \S\ref{ARRAY}) of such. Its type declarator is

\begin{code}
STRUCTURED ARRAY(\emph{field declarations}) OF \emph{type}
\end{code}
where \cod{\emph{field declarations}}, separated by commas, are either \cod{\emph{newid}}s alone or
\cod{\emph{newid}}s followed by (possibly multiple) \cod{ARRAY} dimensions. For example, the declaration
\begin{code}
STRUCTURED ARRAY(a,b,c(1..4,1..2)) OF REAL stra
\end{code}
defines a \cod{STRUCTURE} named \cod{stra} with fields \cod{a}, \cod{b} of type \cod{REAL} and \cod{c} of type
\cod{ARRAY(1..4,1..2) OF REAL}. At the same time, \cod{stra} may also be accessed as an
\cod{ARRAY(0..9) OF REAL}, with automatically determined bounds. This is particularly
useful when the new type has to be subjected to the linear \cod{ARRAY} operations of \S\ref{ARRAY_ops}. For
instance, one can define a ``Vector'' structure with fields x,y,z and then syntetically denote
vector operations on variables of this type.

\subsection{POINTER\label{POINTER}\index{POINTER}}

The type declarator for pointers is \cod{POINTER TO} (prefix) or \cod{\^{}} (postfix)
(see \S\ref{Postfix}). For example

\begin{code}
POINTER TO INTEGER a,b; INTEGER c
\end{code}
and
\begin{code}
INTEGER a\^{},b\^{},c
\end{code}
are equivalent declarations. The notation for pointer dereferencing is
described in \S\ref{Concealed_Pointer_Lookup}.

A null pointer is denoted by the predefined constant

\begin{code}
NULL \emph{type}
\end{code}
which translates to the C constant \cod{NULL} (usually the address 0) cast to type \emph{type} if the latter already is a \cod{POINTER}, or \cod{POINTER TO \emph{type}} otherwise.

\subsubsection{Uncommitted POINTERs}
A pointer can also be declared with reference to a (yet) uncommitted type name,
by the notation

\begin{code}
POINTER TO \emph{newid}
\end{code}

It is assumed that \cod{\emph{newid}} will later be declared as a type identifier. \cod{VARIABLE}s,
\cod{CONSTANT}s, \cod{ARRAY}s and \cod{STRUCTURE} fields of an uncommitted pointer type can be declared,
assigned and compared to each other like all other \cod{POINTER}s but not dereferenced.
They become \cod{POINTER}s in full rights as soon as a declaration for type \cod{\emph{newid}}
is encountered.

Uncommitted pointers acquire a special meaning if used as formal parameters or
function results (in which case no explicit declaration of type \cod{\emph{newid}} shall
follow): a formal parameter of uncommitted pointer type can receive an actual
parameter of any \cod{POINTER} type (except \cod{POINTER TO ARRAY(*)}, which needs special
treatment) as the actual argument, thus allowing the coding of subroutines for the
manipulation of generic pointers. Type checking will still be enforced based
on the equality of type name. Of course, uncommitted pointers of this kind can
never be dereferenced within the body of the subroutine itself.

See also \S\ref{NEW} and \S\ref{DYNAMIC}.

\subsection{DYNAMIC POINTER\label{DYNAMIC}\index{Object-oriented features}}

\CPL\ is not an object-oriented language, in the sense that functions are not
supposed to be declared within object class declarations, and intentionally
so. Object-like behaviour is achieved through overloaded function names (see \S\ref{Coactive_Parameter_Lists}) and implicit type conversion (see \S\ref{STRUCTURE}).

In addition, situations where the type of an object is not fixed at compile
time are catered for by \cod{DYNAMIC} pointers, pointers that can reference and
provide type checking for objects whose type will only be known at run time.
These are declared by the type declarator

\begin{code}
DYNAMIC POINTER
\end{code}
A function call in which such a pointer appears is transparently switched at
run time to the function defined for the type of the actual arguments, or an error is thrown if no such function exists.

Pointers to variables of any type except \cod{ARRAY(*)} and \cod{STORED} can be assigned
to a \cod{DYNAMIC POINTER}. If the scope of assignment is to be restricted, the
declaration becomes

\begin{code}
DYNAMIC POINTER TO \emph{type declarator}
\end{code}

In this case only types which can be implicitly converted to \emph{type declarator}
are acceptable for the variable pointed to.

\subsubsection{Type comparison\label{IS}\index{Type comparison}}

The actual type of the variable pointed to by a dynamic pointer (or, for that
matter, of any variable) can be tested by the \cod{BOOLEAN} expression

\begin{code}
\emph{pointer} IS \emph{type}
\end{code}

For instance, in

\begin{code}
DYNAMIC POINTER dyn

$\cdots$

IF dyn IS INTEGER THEN

$\cdots$

END IF
\end{code}
the \cod{IF} block is executed only if the dynamic pointer \cod{dyn} points at run time to
a variable of \cod{INTEGER} type. In addition, the \cod{IS} test has the side effect of
casting a dynamic pointer to an ordinary pointer of the tested type when the
test succeeds, thus allowing its use as a regular variable. That is, in the
above example \cod{dyn} is implicitly converted to a \cod{POINTER TO INTEGER} within the scope
of the \cod{IF} block. A restricted dynamic pointer can be implicitly
converted to its base type without any test, as compatibility was already enforced at compile time.

\subsection{SUBROUTINE and FUNCTION variables\label{Subroutine_type}\index{Variable call to a subroutine or function}}

A variable containing the address of a subroutine may be declared through the type declarator

\begin{code}
SUBROUTINE( \emph{parameter declarations} )
\end{code}

A variable containing the address of a function may be declared through the type declarator

\begin{code}
FUNCTION( \emph{parameter declarations} )->\emph{type}
\end{code}
or else
\begin{code}
\emph{type} FUNCTION( \emph{parameter declarations} )
\end{code}
the first form being mandatory when the result is of a pointer or compound type and
thus ambiguity can arise.

These type declarators can be used, just like all other type declarators, to
declare variables and formal parameters, or as a part of new compound types. In
addition to being assigned or compared to each other, subroutine variables can
be assigned the name of a compatible subroutine as their value or as the actual
parameter, and can later be called with the same syntax as normal (constant)
subroutines and functions.

\section{Expressions\label{Expressions}\index{Expressions}}

\subsection{INTEGER operators\label{INTEGER_operator}\index{INTEGER operators}}

The infix \cod{INTEGER} arithmetic operators are:	\cod{+}  \cod{-}  \cod{*}  \cod{DIV} and \cod{MOD}.

Notice that the \cod{INTEGER} division operator is denoted by \cod{DIV}, and always produces a rounded-down \cod{INTEGER}
quotient. By contrast \cod{/}  produces a \cod{REAL} result even when the
operands are \cod{INTEGER}s.

\cod{MOD} is the ``modulo'' infix operator. If \cod{a} and \cod{b} are \cod{INTEGER}s,

\begin{code}
a MOD b
\end{code}
represents the remainder of the integer division of \cod{a} by \cod{b}.

\subsubsection{Notice on the truncation of integer division:}
\CPL\ requires that \cod{$a$ MOD $b$} must
always be a positive number, regardless of the sign of \cod{$a/b$}. This is so that
modular arithmetics can be applied to negative as well as to positive numbers,
for example in array index manipulations. Since \cod{$b$*($a$ DIV $b$) + $a$ MOD $b$ = $a$},
consistently \cod{$a$ DIV $b$} must be rounded towards minus infinity, \ie\ \cod{$a$ DIV $b$ =
FLOOR($a/b$)}. Care should be exerted when porting programs from languages that
round up negative division, such as FORTRAN or C.

Operators following the C convention of rounded-towards-zero integer division and signed
modulo are available, if needed, as \cod{CDIV} and \cod{CMOD} respectively.

\subsection{REAL operators\label{REAL_operator}\index{REAL operators}}

The infix \cod{REAL} arithmetic operators are:	\cod{+}  \cod{-}  \cod{*}  \cod{/} and  \cod{\^{}} or \cod{**}.

An infix \cod{\^{}} is the exponentiation operator. (Not to be confused with a postfix \cod{\^{}}, the pointer dereferencing operator or a prefix \cod{\^{}}, the address extraction operator --- see \S\ref{dereferencing}.) \cod{**} is also provided for FORTRAN-accustomed users.

\subsection{Comparison operators\label{Comparison}\index{Comparison operators}}

The infix comparison operators are:	\cod{>}  \cod{>=}  \cod{<}  \cod{<=}  \cod{=}  \cod{\#} and \cod{IS}

All of these produce a \cod{BOOLEAN} result, and have lower precedence than arithmetic operators but higher than boolean operators.

The equality and inequality operators apply to any type, including pointers.
Since pointers are implicitly dereferenced (see \S\ref{Concealed_Pointer_Lookup}),
an ambiguity may arise as to whether the values or addresses of the two sides
must be compared. If explicit dereferencing is not used, the comparison takes place at the earliest
dereferencing level at which types match, but the comparison of two constant
addresses is disallowed (so that \cod{A=B}, where \cod{A} and \cod{B} are two simple variables, denotes the comparison of the values of \cod{A} and \cod{B} rather than their addresses). Explicit type casting or pointer dereferencing may at times be necessary, and is always advised where ambiguities may arise.

The \cod{IS} type-testing operator was described in \S\ref{IS}.

\subsection{BOOLEAN operators\label{BOOLEAN_ops}\index{BOOLEAN operators}}

\subsubsection{AND\label{AND}\index{AND}}

Boolean AND infix operator. Takes precedence over \cod{OR} but not \cod{NOT}.

The \cod{AND} keyword may also appear in a \cod{FOR} clause (see \S\ref{FOR_AND}) and in a \cod{READ} statement (see \S\ref{READ}).

\subsubsection{OR\label{OR}\index{OR}}

Boolean OR infix operator. Has lower precedence than either \cod{AND} or \cod{NOT}.

The \cod{OR} keyword may also appear in a \cod{READ} statement (see \S\ref{READ}).

\subsubsection{NOT\label{NOT}\index{NOT}}

Boolean NOT prefix operator. Has higher precedence than either \cod{AND} or \cod{OR}.

\subsection{Conditional expressions\label{Conditional_expressions}\index{Conditional expressions}}

\begin{code}
IF \emph{boolean} THEN \emph{expr1} ELSE \emph{expr2}
\end{code}

is an expression that takes the value of \cod{\emph{expr1}} if \cod{\emph{boolean}} is true and
\cod{\emph{expr2}} if false. \cod{\emph{expr1}} and \cod{\emph{expr2}} must evaluate to values of the same type
(or be implicitly convertible to the same type) and this is the type of the
result, which can be embedded in a larger expression like any other operator.
Notice that the \cod{ELSE} part is mandatory, as without it the expression's value would be undefined.

The syntax is otherwise very similar to that of the \cod{IF} statement of \S\ref{IF}.

\subsection{Bitwise operators\label{Bitwise}\index{Bitwise}}

The bitwise boolean operators, acting on \cod{INTEGER} operands to produce an \cod{INTEGER} result are denoted as: \cod{BITAND}, \cod{BITOR}, \cod{BITXOR}, \cod{BITNOT}, \cod{RSHIFTED} and \cod{LSHIFTED}.
The C-like notation \cod{\&} \cod{|} \cod{>{}>} \cod{<{}<} is also available for \cod{BITAND} \cod{BITOR} \cod{RSHIFTED} \cod{LSHIFTED}.
\cod{\^{}} and \cod{\textasciitilde} are reserved for other functions (see \S\ref{REAL_operator}, \S\ref{dereferencing} and \S\ref{Assignment}).

\subsection{String concatenation\label{String}\index{String concatenation}}

\cod{STRING} (see \S\ref{STRING}) literals and/or variables written one after the other with no intervening operator (whitespace is
allowed, but newlines must be escaped) are concatenated into a single \cod{STRING}. If all of
the arguments are literals (\ie known at compile time), the result will also be
a literal, otherwise it will be a dynamically allocated \cod{ARRAY(*) OF CHAR} which
is transparently freed at the end of the enclosing code block.

\subsubsection{String conversion}

In addition to \cod{STRING}s, values of other types may also appear in a
concatenation, provided the first item is an actual \cod{STRING} (possibly \cod{""}); they
are then implicitly converted to strings just as if they appeared in a \cod{WRITE} statement (see \S\ref{WRITE}). Please pay attention that a numeric parenthesis that is concatenated
after a string will be interpreted as an \cod{ARRAY} subscript, since the \cod{STRING} is
a legitimate \cod{ARRAY OF CHAR}, unless white space is placed between the string and the
parenthesis. The reverse conversion of a string into a numeric value must be handled explicitly, if needed, through libc String and Array Utilities.

When a string literal is being defined in \cod{<{}<\emph{delimiter}} notation (see \S\ref{STRING}), a
variable included as \cod{\emph{delimiter} \emph{variable} \emph{delimiter}} is concatenated within
the string being defined (whereas \cod{\emph{delimiter} newline} ends the definition).

\subsection{Looping operators\label{Looping_operator}\index{Looping operators}}

The four operators  \cod{SUM}  \cod{PRODUCT}  \cod{MAX}  \cod{MIN} embed a loop inside an expression, to denote respectively
the sum, product, maximum and minimum of their argument over a running index.
\cod{ARGMAX} and \cod{ARGMIN} act like \cod{MAX} and \cod{MIN} but return the value of the index.
They are all constructed as in this instance:

\begin{code}
SUM \emph{expression} FOR \emph{for clause}
\end{code}
and return a value of the same \cod{TYPE} as \cod{\emph{expression}}. The \cod{\emph{for clause}} can be of any one of the
forms that are allowed in a \cod{FOR} loop (see \S\ref{FOR}).

For example:

\begin{code}
WRITE (MAX SIN(x) FOR x=0. TO 2 BY 0.5)+(PRODUCT arr(n) FOR ALL n)
\end{code}

The \cod{MAX} and \cod{MIN} keywords are also used as functions denoting \cod{ARRAY} operations (see \S\ref{MAX}).

\subsection{Builtin functions\label{Builtin}\index{Builtin}}

Most of the following functions can be applied to either \cod{REAL} or \cod{INTEGER} arguments, giving a corresponding result.

\subsubsection{ABS\label{ABS}\index{ABS}}

The \cod{ABS} function returns the absolute value of an \cod{INTEGER}, \cod{REAL} or \cod{COMPLEX} number, or of an \cod{ARRAY} of those.

\subsubsection{CEILING\label{CEILING}\index{CEILING}}

The \cod{CEILING} function returns the larger or equal \cod{INTEGER} to a given \cod{REAL} value.

\subsubsection{FLOOR\label{FLOOR}\index{FLOOR}}

The \cod{FLOOR} function returns the lesser or equal \cod{INTEGER} to a given \cod{REAL} value.

\subsubsection{ROUND\label{ROUND}\index{ROUND}}

The \cod{ROUND} function returns the nearest \cod{INTEGER} to a given \cod{REAL} value.

\subsubsection{MAX\label{MAX}\index{MAX}}

The function

\begin{code}
MAX(\emph{argument} [,\emph{argument}])
\end{code}
accepts any number of scalar arguments and returns their maximum.

The function

\begin{code}
MAX(\emph{array})
\end{code}
returns the maximum of the elements of the \cod{ARRAY} \cod{\emph{array}}.

The function

\begin{code}
ARGMAX(\emph{array})
\end{code}

returns the first index where the maximum is found.

The \cod{MAX} or \cod{ARGMAX} keyword without a following bracket is a looping operator (see \S\ref{Looping_operator}).

\subsubsection{MAXABS\label{MAXABS}\index{MAXABS}}

The function

\begin{code}
MAXABS(\emph{array})
\end{code}
returns the maximum absolute value of the elements of the \cod{ARRAY} \cod{\emph{array}}.

\subsubsection{MIN\label{MIN}\index{MIN}}

The function

\begin{code}
MIN(\emph{argument} [,\emph{argument}])
\end{code}
accepts any number of scalar arguments and returns their minimum.

The function

\begin{code}
MIN(\emph{array})
\end{code}
returns the minimum of the elements of the \cod{ARRAY} \cod{\emph{array}}.

\begin{code}
ARGMIN(\emph{array})
\end{code}

returns the first index where the minimum is found.

The \cod{MIN} or \cod{ARGMIN} keyword without a following bracket is a looping operator (see \S\ref{Looping_operator}).

\subsubsection{NORM\label{NORM}\index{NORM}}

The \cod{NORM} function returns the squared absolute value of either a number or a (uni-- or multi-- dimensional) array.

\subsubsection{ABS}

The function

\begin{code}
ABS(\emph{array})
\end{code}
returns the modulus (square root of the \cod{NORM}) of the \cod{ARRAY} \cod{\emph{array}}.

\subsubsection{Other builtin REAL functions\label{REAL:Builtin}\index{REAL builtin}}

\begin{stabular}{l@{: }l}
\cod{SIN}	&                  sine\\
\cod{COS}	&                  cosine\\
\cod{TAN}	&                  tangent\\
\cod{ATAN}	&                 arctangent\\
\cod{EXP}	&                  exponential\\
\cod{LOG}	&                  logarithm\\
\cod{RAND}	&		uniformly distributed in (0,1) REAL random number\\
\cod{GAUSS}	&		gaussian distributed REAL random number with mean 0 and variance 1\\
\end{stabular}

\subsubsection{Random numbers\label{RAND}\index{Random numbers}}

The \cod{RAND} function takes no argument. It is based on the \cod{INTEGER} C library
function \cod{rand}, whose seed can be changed with \cod{srand} (see the \cod{rand} manpage).

The \cod{GAUSS} function takes no argument. It is based on \cod{RAND} and the analytic formula (see Knuth, The Art of Computer Programming, chapter on random numbers):

\begin{code}
GAUSS = SQRT\{-2*LOG[RAND()]\}*COS[PI*RAND()]
\end{code}

In addition, all C-library mathematical and other functions are
transparently available with their original, generally lowercase, names (see \S\ref{C_interface}).

\subsubsection{Builtin BOOLEAN functions\label{BOOLEAN:builtin}\index{BOOLEAN builtin}}

\begin{stabular}{l@{: }l}
\cod{EOF}	&                  end of file\\
\cod{INPUTREADY}&	test for input ready on a file descriptor or stdin (see \S\ref{INPUTREADY})\\
\cod{ODD}	&                  \cod{INTEGER} argument is odd\\
\cod{READ}	& 	read and report success or failure (see \S\ref{READ})\\
\end{stabular}

\subsubsection{Builtin type-conversion functions\label{type_conversion_builtin}}

\begin{stabular}{l@{: }l}
\cod{CHAR}	&                 \cod{INTEGER} to \cod{CHAR}\\
\cod{INTEGER}	&              \cod{BOOLEAN} to \cod{INTEGER} (\cod{1} for \cod{YES}, \cod{0} for \cod{NO})\\
\cod{INTEGER}	&              \cod{REAL} to \cod{INTEGER} (same as \cod{FLOOR})\\
\cod{REAL}	&                 \cod{INTEGER} or \cod{SINGLE} to \cod{REAL} (may be omitted)\\
\cod{SINGLE}	&               \cod{REAL} to \cod{SINGLE}\\
\end{stabular}

\subsubsection{Command-line parameter access\label{COMMANDLINE}\index{COMMANDLINE}}

The command line through which the program was launched is available in the
predefined \cod{ARRAY(*) OF STRING}

\begin{code}
COMMANDLINE
\end{code}
(the same that is usually named \cod{argv} in C programs). Therefore the first word
passed after the program name is \cod{COMMANDLINE(1)}, the second is \cod{COMMANDLINE(2)},
etc. The program name itself is \cod{COMMANDLINE(0)}. The index of the last word
present, just as for all \cod{ARRAY}s, is \cod{COMMANDLINE.HI}.

\section{Assignment\label{Assignment}\index{Assignment}}

Assignment is performed by a single equals \cod{=} sign. The \lhs\ must be a \cod{POINTER} (which the constant address of a simple variable automatically is), and the \rhs\ is required to be a value of the \cod{TYPE} pointed to or to be implicitly convertible to it (see \S\ref{Concealed_Pointer_Lookup}). Compound variables can be assigned as a whole if types match.

Owing to their all-encompassing role, assignment and equality/inequality tests
are the only situations in which pointer dereferencing must be explicit. The
left side of an assignment will never be implicitly dereferenced, and must be
dereferenced by a postfix \cod{\^{}} if assignment of the value pointed to is desired.
For increased clarity, the type of the object to be assigned may also be
specified, as in the example (after the declaration \cod{REAL a\^{},b}):

	\cod{REAL a=b}	equivalent to      \cod{a\^{}=b}.

At any position in the expression appearing on the right-hand side of an assignment, the symbol \cod{\textasciitilde} can be used as a
place-holder for the \lhs. For example:

\begin{code}
counter = \textasciitilde + 2

listelement = \textasciitilde.next

A(i,j,k) = 3 * LOG(\textasciitilde) + gamma(\textasciitilde)
\end{code}

\subsection{Pointer arithmetics\label{INTO}\index{Pointer arithmetics}}
Pointer arithmetics is the most powerful feature C has in common with assembly
language. It is, however, also one of its weakest points as far as bounds
crossing is concerned. For these reasons \CPL\ allows pointer arithmetics in restricted forms. One of these is subarray  extraction (see \S\ref{subarray}). The other is
the type declarator

\begin{code}
POINTER INTO \emph{array} [, \emph{array}]
\end{code}
which defines a pointer that can be decremented or incremented provided it
stays within the bounds of one of the listed \cod{\emph{array}}s. Arithmetics on such a pointer is
allowed just as if it were an ordinary \cod{INTEGER}; dereferencing is not implicit,
as for ordinary \cod{POINTER}s it would be, but is obtained by using it as an index into one of 
its base arrays. For instance as in:

\begin{code}
REAL R(10)

R(5)=0.5

POINTER INTO R ip

ip=3

ip=ip+2

WRITE R(ip)
\end{code}

In other words, code is written just as if an \cod{INTEGER} index were used (and actually works if just the \cod{POINTER INTO} type declarator is replaced by \cod{INTEGER} without any other change), but the
program is compiled to use a pointer instead. This can sometimes improve
performance, especially in short loops that perform few repeating operations.

\subsection{Dereferencing operator\label{dereferencing}\index{dereferencing}}

Explicitly specifying a dereferencing is optional but never forbidden. When necessary, the dereferencing operator is a postfix \cod{\^{}} (like in Pascal).
Vice versa, a prefix \cod{\^{}} forces taking the address rather than the value of a
variable. This construction too is rarely needed; notice however that
the implicit declaration of a previously undeclared \cod{CONSTANT} (to be introduced in \S\ref{Implicit_declaration}),

\begin{code}
c = var
\end{code}
implicitly declares \cod{c} as a copy of the value of \cod{var}, whereas

\begin{code}
c = \^{}var
\end{code}
declares \cod{c} as a \cod{POINTER} containing the address of \cod{var}.

\subsection{Deferred assignment\label{Deferred_assignment}}

When an assignment is performed by a double equals \cod{==} sign, no statement is
generated in place. The \lhs\ is instead defined to be an alias for the
\rhs, and the assignment will be performed (and error messages issued, if any) at the location where this alias
is first used.

Deferred assignment permits the appearance on the \rhs\ of yet undefined
variables and functions, provided these will be defined before the \lhs\ is actually used.
It also allows compile-time constants to be given symbolic names and still be
recognized as such.

This construction is further extended in the \texttt{symbolic.cpl} library.

\section{Memory allocation\label{NEW}\index{Memory allocation}}

Runtime memory allocation is achieved either through the statement

\begin{code}
NEW \emph{pointer} [,\emph{pointer}]
\end{code}
or the function
\begin{code}
NEW \emph{type}
\end{code}

The \cod{NEW} statement allocates (through the \cod{malloc} system call) the space
needed for a variable of the \cod{TYPE} pointed to by \cod{\emph{pointer}} and sets \cod{\emph{pointer}}
to its address. The \cod{NEW} function returns a \cod{POINTER} to a dynamically
allocated variable of \cod{TYPE} \cod{\emph{type}}. If applied to a \cod{POINTER TO STORED} (see \S\ref{STORED}) or \cod{POINTER TO FILE} (see \S\ref{FILE}) type, \cod{NEW} returns a pointer to a temporary file
which will be automatically destroyed when freed or at program termination.

Dynamically allocated space is retained, independently of subroutine scoping,
until it is returned to the system by the \cod{FREE} statement (see below).

\subsection{Freeing memory or file storage\label{FREE}\index{Freeing memory or file storage}}

\begin{code}
FREE \emph{pointer} [,\emph{pointer}]
\end{code}
releases the memory space or the file pointed to by each \cod{\emph{pointer}}, which must have
been previously obtained from \cod{NEW}, \cod{OPEN} (see \S\ref{OPEN}), or \cod{CREATE} (see \S\ref{CREATE}).
\cod{\emph{pointer}} is subsequently zeroed, so that any later dereferencing
generates a visible runtime exception. If \cod{\emph{pointer}} is a file descriptor,
\cod{FREE}ing it also writes out any previously buffered data.

\section{Input/Output\label{Input_Output}\index{Input/Output}}

Text I/O is performed by \cod{READ} and \cod{WRITE} instructions on
variables of type \cod{FILE}, which must be declared like ordinary variables
and associated with a file before being used for I/O. \cod{FILE} variables can be
implicitly declared if they are \cod{CONSTANT}, and can be part of compound types
like all other variables. Association to a file is obtained through \cod{OPEN} and \cod{CREATE} instructions. A \cod{FILE} variable is translated to the C type \cod{FILE*},
and can be used as such, if necessary, in C-library I/O functions transparently accessed through the
C interface of \S\ref{C_interface}.

Sequential binary I/O is performed by adding the \cod{BINARY} keyword to the
\cod{READ} and \cod{WRITE} instructions. No distinction is assumed to be made between text and binary files at the filesystem level.

Random-access binary I/O is where \CPL\ really differs from other languages; this is described in \S\ref{STORED}. A more traditional, but also lower-level, method of random access is offered by the \cod{POSITION} statement and function of \S\ref{POSITION}.

\subsection{FILE\label{FILE}\index{FILE}}

The
\begin{code}
FILE
\end{code}
type descriptor is translated to C type \cod{FILE*}.
The predefined C file descriptors \cod{stdin}, \cod{stdout}, \cod{stderr}
are also predefined in a \CPL\ program (as are, in fact, most other symbols and
functions in the C \emph{stdio} library according to \S\ref{C_interface}).

The more general construction

\begin{code}
FILE OF \emph{type}
\end{code}
will be introduced in connection with random-access files (see \S\ref{STORED}).
Type \cod{FILE} introduced above is also equivalent to \cod{FILE OF CHAR}.

\subsection{WRITE\label{WRITE}\index{WRITE}}

The
\cod{WRITE} output statement is used as follows:

\begin{code}
WRITE [TO \emph{file}] [BY NAME] [\emph{outputexpr}] [./.]
\end{code}

Here \cod{\emph{file}} can be a \cod{FILE} value, such as returned by  \cod{OPEN}
or \cod{CREATE}, or else a string representing a filename. In the latter case
the file is implicitly \cod{CREATE}d before \cod{WRITE}ing and \cod{FREE}d at the end of the statement.
If \cod{TO \emph{file}} is omitted, \cod{stdout} is implied.
If \cod{\emph{file}} is present, \cod{WRITE TO} may be omitted.

An \cod{\emph{outputexpr}} can be

\begin{code}
\emph{value}[:\emph{format}]
\end{code}
or \cod{,} (a comma). \cod{\emph{value}} can be of any type except \cod{POINTER} (which if present undergoes Concealed Pointer Lookup --- see \S\ref{Concealed_Pointer_Lookup}). Compound types such as \cod{ARRAY} and \cod{STRUCTURE} are printed in a standard format which is understandable to
\cod{READ} (as well as to humans). (If \cod{STRUCTURE} elements are \cod{POINTER}s, however, they are printed as
the string ``POINTER'' and cannot be read back.) If a \cod{SUBROUTINE} named \cod{WRITE} is
defined with first argument of type \cod{FILE} and second argument of a user-defined
\cod{TYPE}, it will be automatically used to print variables of such type. Primitive types are printed
in a default format, unless the numeric part of a C format specifier is
included after a colon.

The default format is initially the C library's default, but may be changed
(down to the end of the enclosing scope) by the \cod{DEFAULTFORMAT} declaration, \eg

\begin{code}
DEFAULTFORMAT 1.15
\end{code}

A \cod{,} in the sequence of \cod{\emph{outputexpr}}s is printed as a tab. \cod{\emph{outputexpr}}s may
also be written one after the other without any comma, in  which case they are
just adjoined without any intervening space. Please pay attention that a
numeric parenthesis that immediately follows a string will be interpreted as
an \cod{ARRAY} subscript, since the \cod{STRING} is a legitimate \cod{ARRAY OF CHAR}, unless
space is placed between the string and the parenthesis.

By default, \cod{WRITE} puts a newline at the end of its sequence of \cod{\emph{outputexpr}}s.
The optional \cod{./.} symbol suppresses this newline and \cod{FLUSH}es the output instead.

\cod{WRITE} with no \cod{\emph{outputexpr}} and no \cod{./.} prints just a newline.

If the optional \cod{BY NAME} qualification is present, \cod{WRITE} prepends each value
with the expression from which it is produced, in the form \cod{\emph{name}=\emph{value}}. If the
expression is a single identifier, this format can be read back by the \cod{READ BY NAME} statement.

\cod{WRITE} syntax is also adopted for the conversion of general types to a string, as described in \S\ref{String}.

\subsection{READ\label{READ}\index{READ}}

The \cod{READ} input statement is used as follows:

\begin{code}
READ [BY NAME] [FROM \emph{file}] [\emph{variable} [\emph{conjunction} \emph{variable}]]
\end{code}
where \cod{\emph{file}} can be a \cod{FILE} value, such as returned by \cod{OPEN}
or \cod{CREATE}, or else a string representing a filename. In the latter case
the file is implicitly opened before \cod{READ} is executed and closed at the end of the statement. If
\cod{FROM \emph{file}} is omitted, \cod{stdin} is assumed.

\cod{\emph{variable}} must be a \cod{VARIABLE} identifier or a \cod{POINTER} to a variable, which can be of any type except \cod{POINTER}. Compound types such as \cod{ARRAY} and \cod{STRUCTURE} are also accepted; they are expected to appear in the same clear-text format as produced by \cod{WRITE}. \cod{\emph{variable}} can also be a literal \cod{STRING}, which is then expected to appear verbatim in the input file. If
\emph{variable} is an \cod{ARRAY(*) OF CHAR}, characters are read until the array is full or
a newline appears. In all other cases, blanks, tabs and newlines are
automatically skipped. Lines starting with an \cod{!} are considered to be comments and skipped.

\cod{\emph{conjunction}} can be one of \cod{,} \cod{AND} or \cod{OR}. If \cod{,} or \cod{AND} is used, the joined variables
must both be compulsorily present in the input or an I/O error will result. If the
conjunction is \cod{OR}, only one or the other is expected to be present and the \cod{READ}
statement succeeds and yields control as soon as one is found.

\cod{READ} with no variable is also allowed and just blocks until a newline is
received, skipping any other characters that precede it. This statement can be
used to skip lines in a file or to create a pause in execution until a newline
character appears in \cod{stdin}.

\cod{READ}, with the same syntax, may also appear as a function in a \cod{BOOLEAN} expression, and should then be
pronounced as the past participle of the verb ``to read''. In this form, \cod{READ} returns a \cod{BOOLEAN} value saying whether reading
was successful or not and never generates an I/O error.

If the optional \cod{BY NAME} qualifier is present, \cod{READ} expects to find the value of
each variable preceded by its name in the input file in the form \cod{\emph{name}=\emph{value}}.

\subsubsection{FROM\label{FROM}\index{FROM}}

As an auxiliary keyword, \cod{FROM} is used in \cod{READ} statements as detailed above.

In addition, the construction

\begin{code}
\emph{type declarator} FROM \emph{file}
\end{code}
represents a value of type \cod{\emph{type declarator}} read from the file descriptor
\cod{\emph{file}}. This value can directly appear in an expression just as the value of a
function can. As an example, the statement

\begin{code}
WRITE 3*(INTEGER FROM stdin)
\end{code}
can replace the sequence of three statements
\begin{code}
INTEGER temp

READ FROM stdin temp

WRITE TO stdout 3*temp
\end{code}

\subsubsection{Console prompting and reading\label{ASK}\index{Console prompting and reading}}

The statement
\begin{code}
ASK [\emph{type}] ["\emph{literal}":] \emph{variable} [, ["\emph{literal}":] \emph{variable}]
\end{code}
has the effect that a prompt is automatically written to \cod{stderr} for each
variable and the value of the variable is subsequently read from \cod{stdin}. The prompt is the optional \cod{\emph{literal}} if present, or the name of the variable followed by a \cod{?} otherwise.

Console prompting can be compounded with \cod{CONSTANT} declaration. If the optional \cod{\emph{type}} is
present, each variable is declared as a \cod{CONSTANT} of this type before being asked.

Under \texttt{icpl}, this console input functionality becomes extended: an \texttt{icpl} shell is opened
whenever the program is waiting for input from its controlling terminal, and any number of \texttt{icpl} commands can be executed
interactively before answering. The first expression given alone on the command line, which \texttt{icpl} would normally print, gets returned as input instead, and the waiting program resumes.
This extension is available to either an interpreted or a compiled program, 
provided the latter is run under \texttt{icpl}.

\subsubsection{Test for input ready on a file descriptor or stdin\label{INPUTREADY}\index{Test for input ready on a file descriptor or stdin}}

The \cod{BOOLEAN} function

\begin{code}
INPUTREADY(\emph{file})
\end{code}
returns \cod{YES} if there are characters available for input on \cod{FILE} variable
\cod{\emph{file}}. This is mainly of use when \cod{\emph{file}} is a terminal, to check whether keys have
been pressed before blocking on a read. Notice that if the terminal is in line-buffered
mode (as unix terminals are by default), characters become available only
after a line feed has been entered. Terminal mode may be altered by commands in \texttt{CHARbyCHAR.cpl}. Other variants:

\begin{code}
INPUTREADY()
\end{code}
is a shorthand for \cod{INPUTREADY(stdin)}.

\begin{code}
INPUTREADY(\emph{file},\emph{time})
\end{code}
waits for input for the given \cod{\emph{time}} (in seconds, a \cod{REAL}) before returning a \cod{NO}.

\begin{code}
INPUTREADY(\emph{time})
\end{code}
is a shorthand for \cod{INPUTREADY(stdin,\emph{time})}.

\subsection{Binary (unformatted) Input/Output\label{BINARY}\index{Binary (unformatted) Input/Output}}

Sequential binary I/O is obtained by either appending or prepending the word \cod{BINARY} to \cod{READ} and \cod{WRITE}, as in the following stencils:

\begin{code}
READ BINARY FROM \emph{file} \emph{variable} [,\emph{variable}]

WRITE BINARY TO \emph{file} \emph{outputexpr} [,\emph{outputexpr}]
\end{code}

In binary I/O no format need, of course, be specified and no newline or tab is
automatically added by \cod{WRITE} statements. The syntax is otherwise unchanged.

\subsubsection{Read or set a file's current position\label{POSITION}\index{Read or set a file's current position}}

As a primitive form of random-access binary I/O, the \cod{POSITION} statement can be used to
get and set the reading and writing position in the file. A more powerful, and
frequently preferable, mechanism is provided by the \cod{STORED} declaration of the next section.

The statement

\begin{code}
POSITION \emph{file}, \emph{n}
\end{code}
sets the current position in file  \cod{\emph{file}} so the next \cod{READ} or \cod{WRITE}
statement will take effect at byte number \cod{\emph{n}}.

The function

\begin{code}
POSITION(\emph{file})
\end{code}
returns the current \cod{READ}/\cod{WRITE} position in file \cod{\emph{file}} as an \cod{INTEGER} result (provided the filesystem-returned position fits in such type).

\subsection{Random-access files as disk-STORED variables\label{STORED}\index{Random-access files as disk-resident variables}}

A random-access binary file typically contains fields and records at fixed
addresses, which can be accessed, as described above, by \cod{POSITION}ing the file cursor before each \cod{READ} or \cod{WRITE} \cod{BINARY}. However, this process requires knowing the numerical address of each field and can be error-prone.

\CPL\ provides a powerful alternative: a pointer to a compound variable residing
in a disk file. Records and fields are then defined by just declaring a
suitable compound type through the usual \cod{STRUCTURE} and \cod{ARRAY}
declarations. For this purpose a type declaration prefixed by the \cod{STORED} keyword is used, as in the example:

\begin{code}
POINTER TO STORED ARRAY(1..10) OF STRUCTURE(REAL x,y) v
\end{code}
which defines a variable named \cod{v} as the pointer to a random-access file
containing 10 records composed of two real numbers each. \cod{v} is in fact a file descriptor, which can be obtained through \cod{OPEN} or \cod{CREATE} just like any other file descriptor, but is also at the same time a \cod{POINTER} and can alternatively be obtained as an anonimous file from \cod{NEW}. Once \cod{v} is assigned a file descriptor, reading and writing to it
is achieved by accessing \cod{v} just as if it were an ordinary \cod{POINTER} to memory. In the above example, since Concealed Pointer Lookup is in effect,

\begin{code}
v(4).x = 3.14
\end{code}
and
\begin{code}
WRITE LOG(v(8).y)
\end{code}
are valid statements. The simple rule is that anything that could be done if \cod{v} had been declared as a normal \cod{POINTER} is still allowed, but the actual storage occurs in a disk file rather than in memory.

Just as for any \cod{POINTER}, accessing a variable of \cod{POINTER TO STORED} type without first assigning a file descriptor to it is an error (caught by 
\texttt{rtchecks.cpl} if active). On the other hand, the declaration
\begin{code}
STORED ARRAY(1..10) OF STRUCTURE(REAL x,y) v
\end{code}
(continuing in the same example) actually reserves storage for the \cod{v} variable, just like the corresponding declaration without the \cod{STORED} keyword would, by
assigning a temporary file to it. This temporary file is
automatically opened and closed as necessary and does not survive the end of the program.

When declared with the special dimension \cod{*} (which can only appear as the leading index), a \cod{STORED ARRAY} can be extended (by just writing to it) to an a-priori undetermined number of records (numbered from 0). The shorthand

\begin{code}
FILE OF \emph{type}
\end{code}
is equivalent to
\begin{code}
POINTER TO STORED ARRAY(*) OF \emph{type}
\end{code}
The \cod{FILE} type declarator is a shorthand for \cod{FILE OF CHAR}.

\cod{STORED ARRAY}s and \cod{POINTER TO STORED ARRAY}s with multiple \cod{*} dimensions are
allowed when used as function formal parameters, just like their non-\cod{STORED} equivalents are.
The corresponding actual parameter must be a \cod{STORED}\cod{ARRAY} (or subarray) of matching number of dimensions and type (not just an \cod{OPEN}ed file).

\subsection{Opening and closing files\label{OPEN}\index{Opening and closing files}}

\subsubsection{OPEN}

The statement

\begin{code}
OPEN \emph{fd}, \emph{filename}
\end{code}
opens file descriptor \cod{\emph{fd}} for the file identified by the string \cod{\emph{filename}} at the filesystem level.
\cod{\emph{fd}} must have been previously declared as one of the \cod{POINTER TO STORED} types.

The function

\begin{code}
OPEN(\emph{filename})
\end{code}
returns a file descriptor of type \cod{FILE} as its value.

With \cod{OPEN}, file \cod{\emph{filename}} is opened for both reading and writing, and is created if it does
not exist, but is never truncated. Therefore, it may be written over but any
data which are not overwritten will remain there. If a possible file of the
same name is to be cleared before writing, \cod{CREATE} should be used
instead.

(\cod{OPEN} calls the \cod{open} system function with options \cod{O\_RDWR|O\_CREAT}. For finer control, the C \cod{open} or \cod{fopen} function may be called directly, just like all other libc functions, as explained in \S\ref{C_interface}.)

\subsubsection{CREATE\label{CREATE}\index{CREATE}}

The statement

\begin{code}
CREATE \emph{fd}, \emph{filename}
\end{code}
opens file descriptor \cod{\emph{fd}} for the file identified by the string
\cod{\emph{filename}} at the filesystem level. \cod{\emph{fd}} must have been previously declared as one of the \cod{POINTER TO STORED} types.

The function

\begin{code}
CREATE(\emph{filename})
\end{code}
returns a file descriptor of type \cod{FILE} as its value.

With \cod{CREATE}, file \cod{\emph{filename}} is opened for both reading and writing, is created if it does not
exist, and is truncated to zero length. Therefore, it may be reread after it
has been written, but any data in a possible pre-existing file of the same name are
lost. If a file of the same name were to be preserved, \cod{OPEN}
should be used instead.

(\cod{CREATE} calls the \cod{open} system function with options \cod{O\_RDWR|O\_CREAT|}\cod{O\_TRUNC}. For finer control over options, the C \cod{open} or \cod{fcreate} function may be called directly, just like all other libc functions, as explained in \S\ref{C_interface}.)

\subsubsection{FLUSH\label{FLUSH}\index{FLUSH}}

The statement

\begin{code}
FLUSH \emph{fd}
\end{code}
writes out any buffered data for file descriptor \cod{\emph{fd}}.

\section{Control statements\label{Control}\index{Control statements}}

\subsection{Conditional execution\label{IF}\index{Conditional execution}}

\cod{IF} statements have a single-line form and a multi-line form. (For the additional
use of \cod{IF} inside expressions see \S\ref{Conditional_expressions}.) Like in BASIC,
single-line \cod{IF} statements terminate at the end of a line, whereas multi-line \cod{IF}
statements are explicitly terminated by \cod{END IF}.  Multi-line \cod{IF}
statements have the following structure:

\begin{code}
IF \emph{boolean} THEN

\  \emph{code block}

END IF
\end{code}
or
\begin{code}
IF \emph{boolean} THEN

\  \emph{code block}

\emph{else statement}
\end{code}

They are recognized by \cod{THEN} being the last word on a line.
Single-line \cod{IF} statements have the structure

\begin{code}
IF \emph{boolean} THEN \emph{single-line block}
\end{code}
or
\begin{code}
IF \emph{boolean} THEN \emph{single-line block} END IF
\end{code}
or
\begin{code}
IF \emph{boolean} THEN \emph{single-line block} \emph{else statement}
\end{code}

A \cod{\emph{single-line block}} is formed of multiple statements separated by semicolons with no intervening newlines. It may, however, contain compound statements (such as a \cod{LOOP} or another \cod{IF} statement) that in turn contain newlines in their own body.

The optional \cod{\emph{else statement}} has the structure

\begin{code}
ELSE

\  \emph{code block}

END IF
\end{code}
or
\begin{code}
ELSE \emph{single-line block}
\end{code}
or
\begin{code}
ELSE \emph{single-line block} END IF
\end{code}
in the multi-line form \cod{ELSE} being followed (with possible intervening whitespace) by a newline character.

Multiple-choice conditionals are simply obtained by nesting another \cod{IF} in the \cod{\emph{else statement}} as follows:

\begin{code}
IF \emph{boolean} THEN

\  \emph{code block}

ELSE IF \emph{other boolean} THEN

\  \emph{other block}

ELSE

\  \emph{yet another block}

END IF

\end{code}

\subsection{Loops\label{LOOP}\index{Loops}}

Loops tested at their beginning (which are to be executed zero or more times) are coded as

\begin{code}
LOOP \emph{test}

\  \emph{code block}

REPEAT [LOOP]
\end{code}
or
\begin{code}
LOOP \emph{name} \emph{test}

\  \emph{code block}

REPEAT \emph{name}
\end{code}

Loops tested at their end (which are to be executed one or more times) are
coded as

\begin{code}
LOOP

\  \emph{code block}

REPEAT [LOOP] \emph{test}
\end{code}
or
\begin{code}
LOOP \emph{name}

\  \emph{code block}

REPEAT \emph{name} \emph{test}
\end{code}
or else by the \cod{DO} short form of \S\ref{DO}.

When a \cod{\emph{name}} is specified, the corresponding \cod{LOOP} can be terminated at any point inside its body (or an inner nested scope) by the 
\begin{code}
EXIT \emph{name}
\end{code}
statement. A named loop without any \cod{\emph{test}} is also allowed, and represents an infinite loop that can only be
terminated by an explicit \cod{EXIT}.

The test \cod{\emph{test}} may be one of \cod{WHILE}, \cod{UNTIL}, \cod{FOR} as described below.

\subsubsection{WHILE\label{WHILE}\index{WHILE loops}}

A simple test that can be used to decide whether to iterate a loop is

\begin{code}
WHILE \emph{boolean}
\end{code}

For instance,

\begin{code}
DO i=i+1 WHILE i<10
\end{code}
will repeat while the indicated boolean condition is true.
\cod{WHILE} is equivalent to \cod{UNTIL NOT}.

\subsubsection{UNTIL\label{UNTIL}\index{UNTIL loops}}

A simple test that can be used to decide whether to terminate a loop is

\begin{code}
UNTIL \emph{boolean}
\end{code}

For instance,

\begin{code}
DO i=i+1 UNTIL i>10
\end{code}

will repeat until the boolean condition becomes true.

\cod{UNTIL} is equivalent to \cod{WHILE NOT}.

\subsection{FOR loops\label{FOR}\index{FOR loops}}

The \cod{\emph{test}} for indexed loops is

\begin{code}
FOR \emph{for clause}
\end{code}

where the \cod{\emph{for clause}} can assume one of several forms. Its basic incarnation is either

\begin{code}
\emph{index} = \emph{lbound} TO \emph{ubound} [BY \emph{step}]
\end{code}
or
\begin{code}
\emph{index} = \emph{ubound} DOWN TO \emph{lbound} [BY \emph{step}]
\end{code}
where \cod{\emph{index}} is incremented by \cod{\emph{step}} in each iteration in the first case and
decremented in the second. The loop terminates when \cod{\emph{index}} exceeds \cod{\emph{ubound}} in
the first case or drops below \cod{\emph{lbound}} in the second. If \cod{BY \emph{step}} is omitted,
\cod{BY 1} is assumed. \cod{\emph{lbound}}, \cod{\emph{ubound}} and \cod{\emph{step}} can all be arbitrary
expressions; notice that a negative \cod{\emph{step}} in either case generates a
never-ending loop.

Both \cod{INTEGER} and \cod{REAL} types are allowed. \cod{\emph{index}} may be an already existing
variable or be implicitly declared by the loop itself (as a particular case of \S\ref{Implicit_declaration}).
In the latter case it acquires the type of the loop bounds and behaves as a
\cod{CONSTANT} within the body of the loop. If the index is \cod{REAL}, \cod{BY} cannot be omitted.

A \cod{\emph{for clause}} can appear in a \cod{LOOP}, a \cod{DO} or one of the looping operators of \S\ref{Looping_operator}. In addition to the two above basic forms, it can have the following
variants:

\subsubsection{AND in a FOR loop\label{FOR_AND}\index{FOR AND}}

Two or more \cod{FOR} loops can be combined in a single \cod{\emph{test}} through the \cod{AND} keyword (not to be confused with the boolean \cod{AND} operator), as in the
following example:

\begin{code}
LOOP FOR i1=l1 TO u1 AND i2=l2 TO u2

\  \emph{code block}

REPEAT
\end{code}
which is equivalent to

\begin{code}
LOOP FOR i1=l1 TO u1

\  LOOP FOR i2=l2 TO u2
  
\    \emph{code block}
    
\  REPEAT
  
REPEAT
\end{code}

\subsubsection{ALL\label{ALL}\index{ALL}}

A \cod{\emph{for clause}} of the following form:

\begin{code}
LOOP FOR ALL \emph{index} [,\emph{index}]
\end{code}
denotes a loop whose bounds are automatically determined as the common bounds of all
\cod{ARRAY}s that \cod{\emph{index}} is used in (while resulting in an error message if such bounds differ from each other). Its use is particularly convenient for short loops and should preferably be limited to such.

An \cod{ALL} clause can also be part of a multiple loop formed with the \cod{AND} keyword, as well as denote a multiple loop itself by containing
multiple indices separated by a comma. In addition, the lower and upper
bounds that would be automatically assigned by \cod{ALL} to the index are also available as \cod{LO}
and \cod{HI} respectively within the scope of a standard \cod{\emph{for clause}}, like in the example:

\begin{code}
DO WRITE A(i,j,k) FOR ALL i,j AND k=LO+2 TO HI-3
\end{code}

\subsubsection{TIMES\label{TIMES}\index{TIMES}}

A statement of the form

\begin{code}
LOOP FOR \emph{number} TIMES
\end{code}
does exactly what it says, and comes handy when just an anonimous counter is needed.

\subsubsection{IN\label{IN}\index{IN}}
Yet another form of \cod{\emph{for clause}} is

\begin{code}
LOOP FOR \emph{element} IN \emph{array}
\end{code}

Here \cod{\emph{element}} represents a \cod{POINTER} to elements of the \cod{ARRAY}
\cod{\emph{array}}. A \cod{LOOP} is performed in which \cod{\emph{element}} runs sequentially through all
the elements of the array (or its subset specified through subarray selection, see \S\ref{subarray}). As a particular case, \cod{\emph{array}} can also be a \cod{CONSTANT} array constructed from a set of elements of the same \cod{TYPE} enclosed in brackets, as in

\begin{code}
LOOP FOR \emph{n} IN (2,5,11)
\end{code}

\subsubsection{EXCEPT\label{EXCEPT}\index{EXCEPT}}

\begin{code}
LOOP FOR \emph{for clause} EXCEPT \emph{condition} ["," \emph{condition}]
\end{code}

The \cod{EXCEPT} qualifier added to any of the \cod{FOR} loop constructions excludes
exceptional values of the index from the iteration. \cod{\emph{condition}} may be either
an expression of \cod{BOOLEAN} type, meaning that the loop is skipped when this condition is \cod{TRUE}, or an expression of the same type as the index, meaning
that the given value is excluded.






\subsection{DO loops (short form)\label{DO}\index{Loops (short form)}}

A short form for nameless loops tested at the end (which are always executed
one or more times) is

\begin{code}
DO \emph{code block} \emph{test}
\end{code}
where \cod{\emph{test}} may be any one of \cod{WHILE} (see \S\ref{WHILE}), \cod{UNTIL} (see \S\ref{UNTIL}), \cod{FOR} (see \S\ref{FOR}).

\subsection{INLINE loops\label{INLINE_LOOP}\index{INLINE LOOP}}

The statement

\begin{code}
INLINE LOOP [\emph{name}] FOR \emph{element} IN \emph{array}

\  \emph{code block}

REPEAT [\emph{name}]
\end{code}
expands in line by repeating the \cod{\emph{code block}} once for each element of \cod{\emph{array}}, whose
dimensions must be known at compile time.

\subsection{Multiple-choice CASE selection\label{CASE}\index{Multiple-choice selection}}

A multiple-choice branch is specified as

\begin{code}
CASE \emph{integer} OF

\emph{tag1}: \emph{block}

\emph{tag2}: \emph{block}

$\cdots$

\emph{tagn}: \emph{block}

[ ELSE \emph{block} ]

END CASE
\end{code}
where \cod{\emph{integer}} is an expression of \cod{INTEGER} type and \cod{\emph{tag1}}, \cod{\emph{tag2}},\ldots \cod{\emph{tagn}} are
\cod{INTEGER} compile-time constants. A list of values separated by commas may also
appear as a tag.

\subsection{EXIT\label{EXIT}\index{EXIT}}

The statement
\begin{code}
EXIT \emph{name}
\end{code}
transfers control past the end of the \cod{SUBROUTINE}, \cod{FUNCTION}, \cod{MODULE} or \cod{LOOP} labelled \cod{\emph{name}}. \cod{EXIT} is allowed to appear in an inner nested scope, such as a further loop or an \cod{IF} statement.

\subsection{END\label{END}\index{END}}

The statement
\begin{code}
END \emph{name}
\end{code}
marks the end of the body of a \cod{SUBROUTINE}, \cod{FUNCTION} or \cod{MODULE} named \cod{\emph{name}}. \cod{END C SECTION}, \cod{END FRI SECTION}, \cod{END IF}, \cod{END CASE}, \cod{END WITH} and \cod{END TRAP}
mark the end of the corresponding statements.

\subsection{STOP\label{STOP}\index{STOP}}

The statement
\begin{code}
STOP
\end{code}
stops the program and exits (unless intercepted by a \cod{TRAP}).

\subsection{ERROR\label{ERROR}\index{Error signal}}

The statement

\begin{code}
ERROR \emph{outputexpr}
\end{code}
writes \cod{\emph{outputexpr}} to the predefined string variable \cod{ERRORMESSAGE} and signals
an error. Unless a \cod{TRAP} has been set, this causes the program to exit
after writing \cod{ERRORMESSAGE} to \cod{stderr}.

\subsubsection{Error TRAP\label{TRAP}\index{Error handling}}

An error, either caused by a system signal or by the \cod{ERROR} statement,
normally causes the program to terminate after writing a message to \cod{stderr}. An error trap may be set by the statement

\begin{code}
TRAP

\  \emph{error handling}

END TRAP
\end{code}
or
\begin{code}
TRAP \emph{literal}

\  \emph{error handling}

END TRAP
\end{code}

In the first form all errors are trapped; in the second, only those whose error
message begins with \cod{\emph{literal}}. During normal execution, the \cod{\emph{error}} \cod{\emph{handling}} block is skipped. At the moment an error is triggered, either by a system signal or by an explicit \cod{ERROR} statement, within the code that follows \cod{END TRAP} up to the end of the enclosing block, the \cod{\emph{error handling}} code block gets executed.
The error message that would be printed is available in this block in the
predefined string \cod{ERRORMESSAGE}.
At the end of the \cod{\emph{error handling}} block, control is passed to the end of the enclosing block,
where also the scope of the \cod{TRAP} terminates like that of all other
declarations. Within the body of the \cod{TRAP}, the exception may be re-raised if
necessary by the statement

\begin{code}
ERROR ERRORMESSAGE
\end{code}

An exception caused by an external \cod{SIGINT} or by the \cod{STOP} statement is accompanied by an empty \cod{ERRORMESSAGE}.

\newpage
\appendix
\section*{Keyword index}
\hypertarget{Keyword_index}{}
\addcontentsline{toc}{section}{Keyword index}

~

\cod{ABS}: absolute value: see \S\ref{ABS}

\cod{ALL}: loop specifier: see \S\ref{ALL}

\cod{AND}: boolean AND: see \S\ref{AND}

\cod{ARGMAX}:   argument for maximum, function or looping operator: see \S\ref{MAX}
\cod{ARGMIN}:         argument for minimum, function or looping operator: see \S\ref{MIN}

\cod{ARRAY}: subscripted arrays: see \S\ref{ARRAY}

\cod{ASK}: console prompting and reading: see \S\ref{ASK}

\cod{ATAN}: arctangent: see \S\ref{Builtin},

\cod{BINARY}: binary format input/output: see \S\ref{BINARY}

\cod{BITAND}: boolean operator: see \S\ref{Bitwise}

\cod{BITNOT}: boolean operator: see \S\ref{Bitwise}

\cod{BITOR}: boolean operator: see \S\ref{Bitwise}

\cod{BITXOR}: boolean operator: see \S\ref{Bitwise}

\cod{BOOLEAN}: primitive type declarator: see \S\ref{BOOLEAN}

\cod{BY}: used in \cod{FOR}, see \S\ref{FOR}, \cod{READ}, see \S\ref{READ}, \cod{WRITE}, see \S\ref{WRITE}

\cod{FOR}: loop specifier: see \S\ref{FOR}

\cod{READ}: input from character files and devices: see \S\ref{READ}

\cod{WRITE}: output to character files and devices: see \S\ref{WRITE}

\cod{C SECTION}: transparent C source code inclusion: see \S\ref{C_SECTION}

\cod{CASE}: multiple-choice selection: see \S\ref{CASE}

\cod{CEILING}: larger or equal integer: see \S\ref{CEILING}

\cod{CHAR}: primitive type declarator: see \S\ref{CHAR}


\cod{COMMANDLINE}: command line parameter access: see \S\ref{COMMANDLINE}

\cod{CONSTANT}: value which cannot be altered later in the program: see \S\ref{CONSTANT}

\cod{COS}: cosine: see \S\ref{Builtin}

\cod{CREATE}: open a file of zero length: see \S\ref{CREATE}

\cod{DEFAULTFORMAT}: default format: see \S\ref{WRITE}

\cod{DIV}: see \S\ref{INTEGER_operator}

\cod{DO}: conditional and sequential loops: see \S\ref{DO}

\cod{DOWN}: Loop with decreasing index: see \S\ref{FOR}

\cod{ELSE}: used in conditionally executed statement, see \S\ref{IF}, or \cod{CASE} statement, see \S\ref{CASE}

\cod{END}: ends a block: see \S\ref{END}

\cod{ENUM}: enumerated type: see \S\ref{ENUM}

\cod{EOF}: test for end of file: see \S\ref{Builtin}

\cod{ERROR}: signal an error and exit or trigger a TRAP: see \S\ref{ERROR}

\cod{ERRORMESSAGE}: error message string: see \S\ref{TRAP}

\cod{EXCEPT}: loop specifier: see \S\ref{EXCEPT}

\cod{EXIT}: exit from loop, module or subroutine: see \S\ref{EXIT}

\cod{EXP}: exponential: see \S\ref{Builtin},

\cod{FALSE}: BOOLEAN value: see \S\ref{BOOLEAN}

\cod{FILE}: serial file-descriptor type: see \S\ref{FILE}

\cod{FLOOR}: smaller or equal integer: see \S\ref{FLOOR}

\cod{FLUSH}: flush a file's write buffer: see \S\ref{FLUSH}

\cod{FOLLOWS}: Function prototypes: see \S\ref{FOLLOWS}

\cod{FOR}: loop specifier: see \S\ref{FOR}

\cod{FORTRANCALL}: see \S\ref{FORTRAN_interface}

\cod{FORTRANFUNCTION}: see \S\ref{FORTRAN_interface}

\cod{FREE}: release dynamically allocated memory or file: see \S\ref{FREE}

\cod{FRI SECTION}: dynamic extension of the language: see \S\ref{FRI_SECTION}

\cod{FROM}: used in READ statements: see \S\ref{FROM}

\cod{FUNCTION}: Function declaration: see \S\ref{FUNCTION}

\cod{GAUSS}: gaussian distributed random REAL number with variance 1: see \S\ref{RAND}

\cod{HI}: upper bound of an array index: see \S\ref{HI}

\cod{HI1}: see \S\ref{HI}

\cod{HI2}: see \S\ref{HI}

\cod{HI3}: see \S\ref{HI}

\cod{IF}: conditionally executed statement: see \S\ref{IF}

\cod{IN}: loop specifier: see \S\ref{IN}

\cod{INCLUDE}: source file inclusion: see \S\ref{INCLUDE}

\cod{INLINE}: Subroutines or functions as macros: see \S\ref{INLINE}

\cod{INPUTREADY}: test for input ready on a file descriptor or stdin: see \S\ref{INPUTREADY}

\cod{INTEGER}: primitive type declarator: see \S\ref{INTEGER}

\cod{INTO}: Pointer arithmetics: see \S\ref{INTO}

\cod{IS}: type comparison operator: see \S\ref{IS}

\cod{LENGTH}: number of elements of an array: see \S\ref{LENGTH}

\cod{LO}: lower bound of an array index: see \S\ref{LO}

\cod{LO1}: see \S\ref{LO}

\cod{LO2}: see \S\ref{LO}

\cod{LO3}: see \S\ref{LO}

\cod{LOG}: logarithm: see \S\ref{Builtin}

\cod{LOOP}: conditional and sequential loops: see \S\ref{LOOP}

\cod{LSHIFTED}: boolean operator: see \S\ref{Bitwise}

\cod{MAX}: maximum value function or looping operator: see \S\ref{MAX}

\cod{MAXABS}: maximum absolute value: see \S\ref{MAXABS}

\cod{MIN}: minimum value function or looping operator: see \S\ref{MIN}

\cod{MOD}: \cod{INTEGER} modulo operator: see \S\ref{INTEGER_operator}

\cod{MODULE}: separately scoped program block: see \S\ref{MODULE}

\cod{NAME}: writing and reading a variable's name: see \S\ref{READ}, \S\ref{WRITE}

\cod{NO}: \cod{BOOLEAN} value: see \S\ref{BOOLEAN}

\cod{NORM}: squared absolute value: see \S\ref{NORM}

\cod{NOT}: boolean NOT: see \S\ref{NOT}

\cod{NULL}: see \S\ref{POINTER}

\cod{DYNAMIC}: object-oriented features: see \S\ref{DYNAMIC}

\cod{ODD}: test for odd number: see \S\ref{Builtin}

\cod{OPEN}: open a file and associate a file descriptor: see \S\ref{OPEN}

\cod{OPTIONAL}: optional function parameters recognized by name: see \S\ref{OPTIONAL}

\cod{OR}: boolean OR: see \S\ref{OR}



\cod{POINTER}: pointer to a memory address: see \S\ref{POINTER}

\cod{POSITION}: read or set a file's current position: see \S\ref{POSITION}

\cod{PRODUCT}: see \S\ref{Looping_operator}

\cod{RAND}: uniformly distributed in (0,1) random REAL number: see \S\ref{RAND}

\cod{READ}: input from character files and devices: see \S\ref{READ}

\cod{REAL}: primitive type declarator: see \S\ref{REAL}

\cod{REPEAT}: see \S\ref{LOOP}

\cod{RESULT}: see \S\ref{FUNCTION}

\cod{RETURN}: see \S\ref{FUNCTION}

\cod{ROUND}: nearest integer: see \S\ref{ROUND}

\cod{RSHIFTED}: boolean operator: see \S\ref{Bitwise}


\cod{SIN}: sine: see \S\ref{Builtin}

\cod{SINGLE}: primitive type declarator: see \S\ref{SINGLE}

\cod{SIZEOF}: memory occupation: see \S\ref{TYPE}

\cod{SQRT}: square root: see \S\ref{Builtin}

\cod{STOP}: stop the program: see \S\ref{STOP}

\cod{STORED}: random-access files as disk-resident variables: see \S\ref{STORED}

\cod{STRIDEOF}: see \S\ref{FORTRAN_interface}

\cod{STRING}: strings of characters: see \S\ref{STRING}

\cod{STRUCTURE}: compound type: see \S\ref{STRUCTURE}

\cod{STRUCTURED ARRAY}: may appear as both a structure and an array: see \S\ref{STRUCTURED}

\cod{SUBROUTINE}: Subroutine declaration: see \S\ref{SUBROUTINE}

\cod{SUM}: see \S\ref{Looping_operator}

\cod{TAN}: tangent: see \S\ref{Builtin}

\cod{THEN}: conditionally executed statement: see \S\ref{IF}

\cod{TIMES}: loop specifier: see \S\ref{TIMES}

\cod{TO}: used in \cod{POINTER}, see\S\ref{POINTER}, \cod{WRITE}, see \S\ref{WRITE}, \cod{FOR}, see \S\ref{FOR}

\cod{WRITE}: output to character files and devices: see \S\ref{WRITE}

\cod{TRANSPOSED}: operation to transpose a matrix: see \S\ref{subarray}

\cod{TRAP}: error handling: see \S\ref{TRAP}

\cod{TRUE}: \cod{BOOLEAN} value: see \S\ref{BOOLEAN}

\cod{TYPE}: declaration of a new type identifier: see \S\ref{TYPE}

\cod{TYPEOF}: extraction pseudo-function: see \S\ref{TYPE}

\cod{UNTIL}: loop terminating condition: see \S\ref{UNTIL}

\cod{USE}: separately compiled modules: see \S\ref{USE}

\cod{VARIABLE}: can be re-assigned multiple times: see \S\ref{VARIABLE}

\cod{WHILE}: loop continuation condition: see \S\ref{WHILE}

\cod{WITH}: implicit access to structure fields and functions: see \S\ref{WITH}

\cod{WRITE}: output to character files and devices: see \S\ref{WRITE}

\cod{YES}: \cod{BOOLEAN} value: see \S\ref{BOOLEAN}











\cod{*}: multiplication operator, or see also \S\ref{ARRAY}

\cod{*2}:        index permutation: see \S\ref{subarray}
\cod{*3}:          index permutation: see \S\ref{subarray}

\cod{**}: see \S\ref{compound_index}; also alternate exponentiation operator

\cod{==}: deferred assignment: see \S\ref{Deferred_assignment}

\cod{\^{}}: exponentiation operator (see \S\ref{REAL_operator}) or \cod{POINTER} dereference (see \S\ref{POINTER})

\cod{\~{}}: place-holder for the l.h.s.: see \S\ref{Assignment}

\cod{<{}*}: insert C code: see \S\ref{C_SECTION}

\cod{<{}<}: multi-line literal: see \S\ref{STRING}

\cod{\#}: unequal comparison operator: see \S\ref{Comparison}

\cod{\#define}: C preprocessor: see \S\ref{C_preprocessor}

\cod{\#else}: C preprocessor: see \S\ref{C_preprocessor}

\cod{\#endif}: C preprocessor: see \S\ref{C_preprocessor}

\cod{\#if}: C preprocessor: see \S\ref{C_preprocessor}

\cod{\#include}: C interface: see \S\ref{C_interface}

\cod{\#includedir}: C interface: see \S\ref{C_interface}

\cod{\#link}: C interface: see \S\ref{C_interface}

\cod{\#undef}: C preprocessor: see \S\ref{C_preprocessor}

\cod{\$}: Einstein convention: see \S\ref{Einstein_convention}

\cod{!}: comment: see \S\ref{Comments}




\printindex
\newpage
\nocite{*}
\hypertarget{References}{}
\bibliography{../CPLreferences/PaoloLuchini,../CPLreferences/MaurizioQuadrio,../CPLreferences/JanPralits,../CPLreferences/DavideGatti,../CPLreferences/MarcoRosti,../CPLreferences/VincenzoCitro}

\begin{thebibliography}{100}

\bibitem{andreolli-quadrio-gatti-2021}
A.~Andreolli, M.~Quadrio, and D.~Gatti.
\newblock Global energy budgets in turbulent {{Couette}} and {{Poiseuille}}
  flows.
\newblock {\em Journal of Fluid Mechanics}, 924, Oct. 2021.
\newblock \href {https://doi.org/10.1017/jfm.2021.598}
  {\path{doi:10.1017/jfm.2021.598}}.

\bibitem{banchetti2020turbulent}
J.~Banchetti, P.~Luchini, and M.~Quadrio.
\newblock Turbulent drag reduction over curved walls.
\newblock {\em Journal of Fluid Mechanics}, 896(A10), 2020.
\newblock \href {https://doi.org/10.1017/jfm.2020.338}
  {\path{doi:10.1017/jfm.2020.338}}.

\bibitem{banchetti-luchini-quadrio-2020}
J.~Banchetti, P.~Luchini, and M.~Quadrio.
\newblock Turbulent drag reduction over curved walls.
\newblock {\em Journal of Fluid Mechanics}, 896, Aug. 2020.

\bibitem{bewley2016methods}
T.~Bewley, P.~Luchini, and J.~Pralits.
\newblock Methods for solution of large optimal control problems that bypass
  open-loop model reduction.
\newblock {\em Meccanica}, 51(12):2997--3014, 2016.
\newblock \href {https://doi.org/10.1007/s11012-016-0547-3}
  {\path{doi:10.1007/s11012-016-0547-3}}.

\bibitem{bewley2007minimal}
T.~Bewley, J.~Pralits, and P.~Luchini.
\newblock Minimal-energy control feedback for stabilization of bluff-body wakes
  based on unstable open-loop eigenvalues and left eigenvectors.
\newblock In {\em Proceedings of the Fifth Conference on Bluff Body Wakes and
  Vortex-Induced Vibrations (BBVIV5)}, pages 129--132, 2007.

\bibitem{Blondeaux2012396}
P.~Blondeaux, J.~Pralits, and G.~Vittori.
\newblock Transition to turbulence at the bottom of a solitary wave.
\newblock {\em Journal of Fluid Mechanics}, 709:396--407, 2012.
\newblock \href {https://doi.org/10.1017/jfm.2012.341}
  {\path{doi:10.1017/jfm.2012.341}}.

\bibitem{Blondeaux2012}
P.~Blondeaux, J.~Pralits, and G.~Vittori.
\newblock Turbulence appearance at the bottom of a solitary wave.
\newblock In {\em Proceedings of the Coastal Engineering Conference}, 2012.
\newblock \href {https://doi.org/10.9753/icce.v33.waves.17}
  {\path{doi:10.9753/icce.v33.waves.17}}.

\bibitem{blondeaux_pralits_vittori_2021}
P.~Blondeaux, J.~O. Pralits, and G.~Vittori.
\newblock On the stability of the boundary layer at the bottom of propagating
  surface waves.
\newblock {\em Journal of Fluid Mechanics}, 928:A26, 2021.
\newblock \href {https://doi.org/10.1017/jfm.2021.807}
  {\path{doi:10.1017/jfm.2021.807}}.

\bibitem{Boi2013}
S.~Boi, A.~Mazzino, and J.~Pralits.
\newblock Minimal model for zero-inertia instabilities in shear-dominated
  non-newtonian flows.
\newblock {\em Physical Review E - Statistical, Nonlinear, and Soft Matter
  Physics}, 88(3), 2013.
\newblock \href {https://doi.org/10.1103/PhysRevE.88.033007}
  {\path{doi:10.1103/PhysRevE.88.033007}}.

\bibitem{Boi2015}
S.~Boi, A.~Mazzino, and J.~Pralits.
\newblock Zero-inertia instabilities in rheopectic fluids.
\newblock In {\em Proceedings - 15th European Turbulence Conference, ETC 2015},
  2015.

\bibitem{bottaro1996linear}
A.~Bottaro and P.~Luchini.
\newblock The linear stability of {G\"ortler} vortices revisited.
\newblock In {\em Mathematical Modeling and Simulation in Hydrodynamic
  Stability}, pages 1--14. World Scientific, 1996.

\bibitem{bottaro1999gortler}
A.~Bottaro and P.~Luchini.
\newblock {G\"ortler} vortices: are they amenable to local eigenvalue analysis?
\newblock {\em European Journal of Mechanics-B/Fluids}, 18(1):47--65, 1999.
\newblock \href {https://doi.org/10.1016/s0997-7546(99)80005-3}
  {\path{doi:10.1016/s0997-7546(99)80005-3}}.

\bibitem{bystrom2007optimal}
M.~G. Bystr{\"o}m, J.~O. Pralits, A.~Hanifi, P.~Luchini, and D.~Henningson.
\newblock Optimal disturbances in three-dimensional boundary-layer flows.
\newblock In {\em 6th ERCOFTAC SIG 33 workshop, Laminar-Turbulent Transition
  Mechanisms, Prediction and Control. June 17-20, 2007, Kleinwalsertal,
  Austria.}, 2007.

\bibitem{carini2015feedback}
M.~Carini, J.~Pralits, and P.~Luchini.
\newblock Feedback control of vortex shedding using a full-order optimal
  compensator.
\newblock {\em Journal of Fluids and Structures}, 53:15--25, 2015.
\newblock \href {https://doi.org/10.1016/j.jfluidstructs.2014.11.011}
  {\path{doi:10.1016/j.jfluidstructs.2014.11.011}}.

\bibitem{carini2013cylinder}
M.~Carini, J.~O. Pralits, and P.~Luchini.
\newblock Cylinder wake stabilization using a minimal energy compensator.
\newblock In {\em ERCOFTAC international symposium on Advances in
  fluid-structure interaction, Mykonos, Greece, June 17-21, 2013}, pages
  335--348, 2016.
\newblock \href {https://doi.org/10.1007/978-3-319-27386-0\_21}
  {\path{doi:10.1007/978-3-319-27386-0\_21}}.

\bibitem{carini2010direct}
M.~Carini and M.~Quadrio.
\newblock Direct-numerical-simulation-based measurement of the mean impulse
  response of homogeneous isotropic turbulence.
\newblock {\em Physical Review E}, 82(6):066301, 2010.
\newblock \href {https://doi.org/10.1103/PhysRevE.82.066301}
  {\path{doi:10.1103/PhysRevE.82.066301}}.

\bibitem{cathalifaud2000algebraic}
P.~Cathalifaud and P.~Luchini.
\newblock Algebraic growth in boundary layers: optimal control by blowing and
  suction at the wall.
\newblock {\em European Journal of Mechanics-B/Fluids}, 19(4):469--490, 2000.
\newblock \href {https://doi.org/10.1016/S0997-7546(00)00128-X}
  {\path{doi:10.1016/S0997-7546(00)00128-X}}.

\bibitem{cathalifaud2000optimal}
P.~Cathalifaud and P.~Luchini.
\newblock Optimal control by blowing and suction at the wall of algebraically
  growing boundary layer disturbances.
\newblock In {\em Laminar-Turbulent Transition}, pages 307--312. Springer,
  Berlin, Heidelberg, 2000.

\bibitem{chiarini-etal-2021}
A.~Chiarini, M.~Mauriello, D.~Gatti, and M.~Quadrio.
\newblock Ascending-descending and direct-inverse cascades of {{Reynolds}}
  stresses in turbulent {{Couette}} flow.
\newblock {\em J. Fluid Mech.}, 930:A9--22, 2021.
\newblock \href {https://doi.org/https://doi.org/10.1017/jfm.2021.886}
  {\path{doi:https://doi.org/10.1017/jfm.2021.886}}.

\bibitem{chiarini-quadrio-2020}
A.~Chiarini and M.~Quadrio.
\newblock The light/dark cycle of microalgae in a thin-layer photobioreactor.
\newblock {\em Journal of Applied Phycology}, 33:183--195, Nov. 2020.
\newblock \href {https://doi.org/10.1007/s10811-020-02310-1}
  {\path{doi:10.1007/s10811-020-02310-1}}.

\bibitem{chiarini-quadrio-2021b}
A.~Chiarini and M.~Quadrio.
\newblock The importance of corner sharpness in the {{BARC}} test case: A
  numerical study.
\newblock {\em Wind \& Structures: an International Journal. Arxiv
  physics.flu-dyn/2109.03522v1}, In press, 2021.
\newblock \href {http://arxiv.org/abs/2109.03522v1}
  {\path{arXiv:2109.03522v1}}.

\bibitem{chiarini-quadrio-2021}
A.~Chiarini and M.~Quadrio.
\newblock The {{Turbulent Flow}} over the {{BARC Rectangular Cylinder}}: A
  {{DNS Study}}.
\newblock {\em Flow, Turbulence and Combustion}, pages 1--25, 2021.
\newblock \href {https://doi.org/10.1007/s10494-021-00254-1}
  {\path{doi:10.1007/s10494-021-00254-1}}.

\bibitem{cimarelli2013prediction}
A.~Cimarelli, B.~Frohnapfel, Y.~Hasegawa, E.~De~Angelis, and M.~Quadrio.
\newblock Prediction of turbulence control for arbitrary periodic spanwise wall
  movement.
\newblock {\em Physics of Fluids}, 25(7):075102, 2013.
\newblock \href {https://doi.org/10.1063/1.4813807}
  {\path{doi:10.1063/1.4813807}}.

\bibitem{citro2015linear}
V.~Citro, F.~Giannetti, L.~Brandt, and P.~Luchini.
\newblock Linear three-dimensional global and asymptotic stability analysis of
  incompressible open cavity flow.
\newblock {\em Journal of Fluid Mechanics}, 768:113--140, 2015.
\newblock \href {https://doi.org/10.1017/jfm.2015.72}
  {\path{doi:10.1017/jfm.2015.72}}.

\bibitem{citro2013unsteady}
V.~Citro and P.~Luchini.
\newblock Unsteady boundary-layer transition prediction.
\newblock In {\em Memorie del XXI Congresso AIMETA 2013, Torino, 17-20 Sep},
  pages 1--9, 2013.

\bibitem{citro2015multiple}
V.~Citro and P.~Luchini.
\newblock Multiple-scale approximation of instabilities in unsteady boundary
  layers.
\newblock {\em European Journal of Mechanics-B/Fluids}, 50:1--8, 2015.
\newblock \href {https://doi.org/10.1016/j.euromechflu.2014.10.004}
  {\path{doi:10.1016/j.euromechflu.2014.10.004}}.

\bibitem{citro2017efficient}
V.~Citro, P.~Luchini, F.~Giannetti, and F.~Auteri.
\newblock Efficient stabilization and acceleration of numerical simulation of
  fluid flows by residual recombination.
\newblock {\em Journal of Computational Physics}, 344:234--246, 2017.
\newblock \href {https://doi.org/10.1016/j.jcp.2017.04.081}
  {\path{doi:10.1016/j.jcp.2017.04.081}}.

\bibitem{coleman2017direct}
G.~Coleman, S.~Pirozzoli, M.~Quadrio, and P.~Spalart.
\newblock Direct numerical simulation and theory of a wall-bounded flow with
  zero skin friction.
\newblock {\em Flow, turbulence and combustion}, 99(3-4):553--564, 2017.
\newblock \href {https://doi.org/10.1007/s10494-017-9834-x}
  {\path{doi:10.1007/s10494-017-9834-x}}.

\bibitem{cormier2016interaction}
M.~Cormier, D.~Gatti, and B.~Frohnapfel.
\newblock Interaction between inner and outer layer in drag-reduced turbulent
  flows.
\newblock {\em PAMM}, 16(1):633--634, 2016.
\newblock \href {https://doi.org/10.1002/pamm.201610305}
  {\path{doi:10.1002/pamm.201610305}}.

\bibitem{de1995application}
P.~De~Matteis, R.~Donelli, and P.~Luchini.
\newblock Application of the ray-tracing theory to the stability analysis of
  three-dimensional incompressible boundary layers.
\newblock In {\em XIII AIDAA Conference}, 1995.

\bibitem{donelli2011global}
R.~Donelli, F.~Giannetti, and P.~Luchini.
\newblock Global stability analysis of open cavity flows in the acoustic limit.
\newblock In {\em XX Congresso Associazione Italiana di Meccanica Teorica e
  Applicata, Bologna 12-15 Sep. 2011}, page~47. Publi\&Stampa, Bologna, 2011.

\bibitem{frohnapfel.etal-2012-Moneytime}
B.~Frohnapfel, Y.~Hasegawa, and M.~Quadrio.
\newblock Money versus time: Evaluation of flow control in terms of energy
  consumption and convenience.
\newblock {\em Journal of Fluid Mechanics}, 700:406--418, 2012.
\newblock \href {https://doi.org/10.1017/jfm.2012.139}
  {\path{doi:10.1017/jfm.2012.139}}.

\bibitem{galantucci2011turbulent}
L.~Galantucci, C.~Barenghi, M.~Sciacca, M.~Quadrio, and P.~Luchini.
\newblock Turbulent superfluid profiles in a counterflow channel.
\newblock {\em Journal of Low Temperature Physics}, 162(3-4):354--360, 2011.
\newblock \href {https://doi.org/10.1007/s10909-010-0266-4}
  {\path{doi:10.1007/s10909-010-0266-4}}.

\bibitem{galantucci2010very}
L.~Galantucci and M.~Quadrio.
\newblock Very fine near-wall structures in turbulent scalar mixing.
\newblock {\em International journal of heat and fluid flow}, 31(4):499--506,
  2010.
\newblock \href {https://doi.org/10.1016/j.ijheatfluidflow.2010.04.002}
  {\path{doi:10.1016/j.ijheatfluidflow.2010.04.002}}.

\bibitem{galantucci2009superfluid}
L.~Galantucci, M.~Quadrio, and P.~Luchini.
\newblock Superfluid vortices in a wall-bounded flow.
\newblock In {\em XIX Congresso AIMETA di Meccanica Teorica e Applicata, Ancona
  14-17 Sep. 2009}, pages 1--10. Aras Edizioni, 2009.

\bibitem{gatti.etal-2020-Structurefunction}
D.~Gatti, A.~Chiarini, A.~Cimarelli, and M.~Quadrio.
\newblock Structure function tensor equations in inhomogeneous turbulence.
\newblock {\em Journal of Fluid Mechanics}, 898, Sept. 2020.
\newblock \href {https://doi.org/10.1017/jfm.2020.399}
  {\path{doi:10.1017/jfm.2020.399}}.

\bibitem{gatti2018global}
D.~Gatti, A.~Cimarelli, Y.~Hasegawa, B.~Frohnapfel, and M.~Quadrio.
\newblock Global energy fluxes in turbulent channels with flow control.
\newblock {\em Journal of Fluid Mechanics}, 857:345--373, 2018.
\newblock \href {https://doi.org/10.1017/jfm.2020.399}
  {\path{doi:10.1017/jfm.2020.399}}.

\bibitem{davide2015experimental}
D.~Gatti, A.~G{\"u}ttler, B.~Frohnapfel, and C.~Tropea.
\newblock Experimental assessment of spanwise-oscillating dielectric
  electroactive surfaces for turbulent drag reduction in an air channel flow.
\newblock {\em Experiment in Fluids}, 56(110):1--15, 2015.
\newblock \href {https://doi.org/10.1007/s00348-015-1983-x}
  {\path{doi:10.1007/s00348-015-1983-x}}.

\bibitem{gatti2013performance}
D.~Gatti and M.~Quadrio.
\newblock Performance losses of drag-reducing spanwise forcing at moderate
  values of the {{Reynolds}} number.
\newblock {\em Physics of Fluids}, 25(12):125109, 2013.
\newblock \href {https://doi.org/10.1017/jfm.2016.485}
  {\path{doi:10.1017/jfm.2016.485}}.

\bibitem{gatti2016reynolds}
D.~Gatti and M.~Quadrio.
\newblock Reynolds-number dependence of turbulent skin-friction drag reduction
  induced by spanwise forcing.
\newblock {\em Journal of Fluid Mechanics}, 802:553--582, 2016.
\newblock \href {https://doi.org/10.1017/jfm.2016.485}
  {\path{doi:10.1017/jfm.2016.485}}.

\bibitem{gatti2018predicting}
D.~Gatti, A.~Stroh, B.~Frohnapfel, and Y.~Hasegawa.
\newblock Predicting turbulent spectra in drag-reduced flows.
\newblock {\em Flow, Turbulence and Combustion}, 100(4):1081--1099, 2018.
\newblock \href {https://doi.org/10.1007/s10494-018-9920-8}
  {\path{doi:10.1007/s10494-018-9920-8}}.

\bibitem{giannetti2019sensitivity}
F.~Giannetti, S.~Camarri, and V.~Citro.
\newblock Sensitivity analysis and passive control of the secondary instability
  in the wake of a cylinder.
\newblock {\em Journal of Fluid Mechanics}, 864:45--72, 2019.
\newblock \href {https://doi.org/https://doi.org/10.1017/jfm.2019.17}
  {\path{doi:https://doi.org/10.1017/jfm.2019.17}}.

\bibitem{giannetti2010structural}
F.~Giannetti, S.~Camarri, and P.~Luchini.
\newblock Structural sensitivity of the secondary instability in the wake of a
  circular cylinder.
\newblock {\em Journal of Fluid Mechanics}, 651:319--337, 2010.
\newblock \href {https://doi.org/10.1017/S0022112009993946}
  {\path{doi:10.1017/S0022112009993946}}.

\bibitem{giannetti2013three}
F.~Giannetti, V.~Citro, L.~Brandt, and P.~Luchini.
\newblock Three-dimensional instability in open cavity flows.
\newblock In {\em XXI Congresso dell'Associazione Italiana di Meccanica Teorica
  ed Applicata (AIMETA)}, pages 1--10, 2013.

\bibitem{giannetti2003receptivity}
F.~Giannetti and P.~Luchini.
\newblock Receptivity of the circular cylinder's first instability.
\newblock In {\em Proc. 5th Eur. Fluid Mech. Conf., Toulouse}, 2003.

\bibitem{giannetti2006leading}
F.~Giannetti and P.~Luchini.
\newblock Leading-edge receptivity by adjoint methods.
\newblock {\em Journal of Fluid Mechanics}, 547:21, 2006.
\newblock \href {https://doi.org/10.1017/S002211200500649X}
  {\path{doi:10.1017/S002211200500649X}}.

\bibitem{giannetti2007structural}
F.~Giannetti and P.~Luchini.
\newblock Structural sensitivity of the first instability of the cylinder wake.
\newblock {\em Journal of Fluid Mechanics}, 581(1):167--197, 2007.
\newblock \href {https://doi.org/10.1017/S0022112007005654}
  {\path{doi:10.1017/S0022112007005654}}.

\bibitem{giannetti2009linear}
F.~Giannetti, P.~Luchini, and L.~Marino.
\newblock Linear stability analysis of three-dimensional lid-driven cavity
  flow.
\newblock In {\em Atti del XIX Congresso AIMETA di Meccanica Teorica e
  Applicata}, pages 14--17. Aras Edizioni Ancona, Italy, 2009.

\bibitem{giannetti2010characterization}
F.~Giannetti, P.~Luchini, and L.~Marino.
\newblock Characterization of the three-dimensional instability in a lid-driven
  cavity by an adjoint based analysis.
\newblock In {\em Seventh IUTAM Symposium on Laminar-Turbulent Transition},
  pages 165--170. Springer, Dordrecht, 2010.
\newblock \href {https://doi.org/10.1007/978-90-481-3723-7-25}
  {\path{doi:10.1007/978-90-481-3723-7-25}}.

\bibitem{giannetti2011stability}
F.~Giannetti, P.~Luchini, and L.~Marino.
\newblock Stability and sensitivity analysis of non-newtonian flow through an
  axisymmetric expansion.
\newblock {\em J Physics: Conference Series}, 318(3):032015, 2011.
\newblock \href {https://doi.org/10.1088/1742-6596/318/3/032015}
  {\path{doi:10.1088/1742-6596/318/3/032015}}.

\bibitem{grea2005ray}
B.-J. {Gr\'ea}, P.~Luchini, and A.~Bottaro.
\newblock Ray theory of flow instability and the formation of caustics in
  boundary layers.
\newblock Technical report, IMFT Internal Report, 2005.

\bibitem{haller2020objective}
G.~Haller, S.~Katsanoulis, M.~Holzner, B.~Frohnapfel, and D.~Gatti.
\newblock Objective barriers to the transport of dynamically active vector
  fields.
\newblock {\em Journal of Fluid Mechanics}, 905, 2020.
\newblock \href {https://doi.org/10.1017/jfm.2020.737}
  {\path{doi:10.1017/jfm.2020.737}}.

\bibitem{hasegawa.etal-2014-Numericalsimulation}
Y.~Hasegawa, M.~Quadrio, and B.~Frohnapfel.
\newblock Numerical simulation of turbulent duct flows at constant power input.
\newblock {\em Journal of Fluid Mechanics}, 750:191--209, 2014.
\newblock \href {https://doi.org/10.1017/jfm.2014.269}
  {\path{doi:10.1017/jfm.2014.269}}.

\bibitem{Isakova2014b}
K.~Isakova, J.~Pralits, R.~Repetto, and M.~Romano.
\newblock Mechanical models of the dynamics of vitreous substitutes.
\newblock {\em BioMed Research International}, 2014, 2014.
\newblock \href {https://doi.org/10.1155/2014/672926}
  {\path{doi:10.1155/2014/672926}}.

\bibitem{Isakova2014a}
K.~Isakova, J.~Pralits, R.~Repetto, and M.~Romano.
\newblock A model for the linear stability of the interface between aqueous
  humor and vitreous substitutes after vitreoretinal surgery.
\newblock {\em Physics of Fluids}, 26(12), 2014.
\newblock \href {https://doi.org/10.1063/1.4902163}
  {\path{doi:10.1063/1.4902163}}.

\bibitem{kaiser2020stages}
F.~Kaiser, B.~Frohnapfel, R.~Ostilla-M{\'o}nico, J.~Kriegseis, D.~E. Rival, and
  D.~Gatti.
\newblock On the stages of vortex decay in an impulsively stopped, rotating
  cylinder.
\newblock {\em Journal of Fluid Mechanics}, 885, 2020.
\newblock \href {https://doi.org/10.1017/jfm.2019.974}
  {\path{doi:10.1017/jfm.2019.974}}.

\bibitem{luchini1994end}
P.~Luchini.
\newblock End-correction integration formulae with optimized terminal sampling
  points.
\newblock {\em Computer physics communications}, 83(2-3):236--244, 1994.
\newblock \href {https://doi.org/10.1016/0010-4655(94)90051-5}
  {\path{doi:10.1016/0010-4655(94)90051-5}}.

\bibitem{luchini1994fourier}
P.~Luchini.
\newblock Fourier analysis of numerical integration formulae.
\newblock {\em Computer physics communications}, 83(2-3):227--235, 1994.
\newblock \href {https://doi.org/10.1016/0010-4655(94)90050-7}
  {\path{doi:10.1016/0010-4655(94)90050-7}}.

\bibitem{luchini1996reynolds}
P.~Luchini.
\newblock Reynolds-number-independent instability of the boundary layer over a
  flat surface.
\newblock {\em Journal of Fluid Mechanics}, 327:101--115, 1996.
\newblock \href {https://doi.org/10.1017/S0022112096008476}
  {\path{doi:10.1017/S0022112096008476}}.

\bibitem{luchini1997computation}
P.~Luchini.
\newblock Computation of three-dimensional {Stokes} flow over complicated
  surfaces ({3D} riblets) using a boundary-independent grid and local
  corrections.
\newblock In {\em 10th European Drag Reduction Meeting, Berlin.}, 1997.

\bibitem{luchini2000reynolds}
P.~Luchini.
\newblock Reynolds-number-independent instability of the boundary layer over a
  flat surface: optimal perturbations.
\newblock {\em Journal of Fluid Mechanics}, 404:289--309, 2000.
\newblock \href {https://doi.org/10.1017/S0022112099007259}
  {\path{doi:10.1017/S0022112099007259}}.

\bibitem{luchini2008acoustic}
P.~Luchini.
\newblock Acoustic streaming and lower-than-laminar drag in controlled channel
  flow.
\newblock In {\em Progress in Industrial Mathematics at ECMI 2006}, pages
  169--177. Springer, Berlin, Heidelberg, 2008.
\newblock \href {https://doi.org/10.1007/978-3-540-71992-2\_12}
  {\path{doi:10.1007/978-3-540-71992-2\_12}}.

\bibitem{luchini2008phase}
P.~Luchini.
\newblock Phase-locked linear response and the optimal feedback control of
  near-wall turbulence.
\newblock {\em Mathematical Physics Models and Engineering Sciences, Liguori
  Editore, Naples}, 2008.

\bibitem{luchini2008receptivity}
P.~Luchini.
\newblock Receptivity to molecular agitation in boundary-layer transition.
\newblock {\em Bull. Am. Phys. Soc.}, 61:HD--005, 2008.

\bibitem{luchini2008role}
P.~Luchini.
\newblock The role of microscopic fluctuations in transition prediction.
\newblock 2008.
\newblock \href {http://arxiv.org/abs/0804.2067} {\path{arXiv:0804.2067}}.

\bibitem{luchini2010thermodynamic}
P.~Luchini.
\newblock A thermodynamic lower bound on transition-triggering disturbances.
\newblock In {\em Seventh IUTAM Symposium on Laminar-Turbulent Transition},
  pages 11--18. Springer, Dordrecht, 2010.
\newblock \href {https://doi.org/10.1007/978-90-481-3723-7-2}
  {\path{doi:10.1007/978-90-481-3723-7-2}}.

\bibitem{luchini2012linearized}
P.~Luchini.
\newblock Linearized boundary conditions at a rough surface.
\newblock {\em Bulletin of the American Physical Society}, 57, 2012.

\bibitem{luchini2013linearized}
P.~Luchini.
\newblock Linearized no-slip boundary conditions at a rough surface.
\newblock {\em Journal of fluid mechanics}, 737:349--367, 2013.
\newblock \href {https://doi.org/10.1017/jfm.2013.574}
  {\path{doi:10.1017/jfm.2013.574}}.

\bibitem{luchini2014receptivity}
P.~Luchini.
\newblock Receptivity to thermal noise in real airfoil configurations.
\newblock {\em Bull. Am. Phys. Soc.}, pages A21--003, 2014.

\bibitem{luchini2015relevance}
P.~Luchini.
\newblock The relevance of longitudinal and transverse protrusion heights for
  drag reduction by a superhydrophobic surface.
\newblock In {\em European Drag Reduction and Flow Control Meeting 2015}, 2015.

\bibitem{luchini2016contradictions}
P.~Luchini.
\newblock Contradictions in the large-wavelength approximation of turbulent
  flow past a wavy bottom.
\newblock In {\em Progress in Turbulence VI}, pages 155--159. Springer, Cham,
  2016.
\newblock \href {https://doi.org/10.1007/978-3-319-29130-7\_28}
  {\path{doi:10.1007/978-3-319-29130-7\_28}}.

\bibitem{luchini2016immersed}
P.~Luchini.
\newblock Immersed-boundary simulations of turbulent flow past a sinusoidally
  undulated river bottom.
\newblock {\em European Journal of Mechanics-B/Fluids}, 55:340--347, 2016.
\newblock \href {https://doi.org/10.1016/j.euromechflu.2015.08.007}
  {\path{doi:10.1016/j.euromechflu.2015.08.007}}.

\bibitem{luchini2016surprising}
P.~Luchini.
\newblock Surprising behaviour in the large-wavelength approximation of
  turbulent flow past a wavy bottom.
\newblock In {\em International Symposium on Stratified Flows}, volume~1, 2016.

\bibitem{luchini2017addendum}
P.~Luchini.
\newblock Addendum to ``{Immersed}-boundary simulations of turbulent flow past
  a sinusoidally undulated river bottom''{[}{Eur.} {J}. {Mech.} {B} {Fluids} 55
  (2016) 340--347{]}.
\newblock {\em European Journal of Mechanics-B/Fluids}, 62:57--58, 2017.
\newblock \href {https://doi.org/10.1016/j.euromechflu.2016.11.013}
  {\path{doi:10.1016/j.euromechflu.2016.11.013}}.

\bibitem{luchini2017receptivity}
P.~Luchini.
\newblock Receptivity to thermal noise of the boundary layer over a swept wing.
\newblock {\em AIAA Journal}, 55(1):121--130, 2017.
\newblock \href {https://doi.org/10.2514/1.J054891}
  {\path{doi:10.2514/1.J054891}}.

\bibitem{luchini2018surprising}
P.~Luchini.
\newblock Surprising behaviour and singularity in the {Saint}-{Venant}
  approximation for a fluid.
\newblock {\em Istituto Lombardo-Accademia di Scienze e Lettere-Incontri di
  Studio}, 2018.

\bibitem{luchini2010methods}
P.~Luchini and T.~Bewley.
\newblock Methods for the solution of very large flow-control problems that
  bypass open-loop model reduction.
\newblock {\em Bull. Am. Phys. Soc.}, 63:AJ--003, 2010.

\bibitem{luchini2006wiener}
P.~Luchini, T.~Bewley, and M.~Quadrio.
\newblock Wiener filters in active-feedback drag reduction of turbulent channel
  flow.
\newblock In {\em 6th EUROMECH Fluid Mechanics Conference (EFMC6)}, 2006.

\bibitem{luchini1996time}
P.~Luchini and A.~Bottaro.
\newblock A time-reversed approach to the study of {G\"ortler} instabilities.
\newblock In {\em Advances in Turbulence VI}, pages 369--370. Springer,
  Dordrecht, 1996.

\bibitem{luchini1998gortler}
P.~Luchini and A.~Bottaro.
\newblock {G\"ortler} vortices: a backward-in-time approach to the receptivity
  problem.
\newblock {\em Journal of Fluid Mechanics}, 363:1--23, 1998.
\newblock \href {https://doi.org/10.1017/S0022112098008970}
  {\path{doi:10.1017/S0022112098008970}}.

\bibitem{luchini2001linear}
P.~Luchini and A.~Bottaro.
\newblock Linear stability and receptivity analyses of the {Stokes} layer
  produced by an impulsively started plate.
\newblock {\em Physics of Fluids}, 13(6):1668--1678, 2001.
\newblock \href {https://doi.org/10.1063/1.1369605}
  {\path{doi:10.1063/1.1369605}}.

\bibitem{luchini2014adjoint}
P.~Luchini and A.~Bottaro.
\newblock Adjoint equations in stability analysis.
\newblock {\em Annual Review of Fluid Mechanics}, 46:493--517, 2014.
\newblock \href {https://doi.org/10.1146/annurev-fluid-010313-141253}
  {\path{doi:10.1146/annurev-fluid-010313-141253}}.

\bibitem{luchini2014direct}
P.~Luchini and A.~Bottaro.
\newblock Direct numerical simulation of flow past superhydrophobic surfaces.
\newblock {\em Bull. Am. Phys. Soc.}, pages E11--004, 2014.

\bibitem{luchini2001optimal}
P.~Luchini, A.~Bottaro, and S.~Zuccher.
\newblock Optimal perturbations and control of nonlinear boundary layer.
\newblock {\em Bull. Am. Phys. Soc.}, 54:JN--007, 2001.

\bibitem{luchini2005acoustic}
P.~Luchini and F.~Charru.
\newblock Acoustic streaming past a vibrating wall.
\newblock {\em Physics of Fluids}, 17(12):122106, 2005.
\newblock \href {https://doi.org/10.1063/1.2149314}
  {\path{doi:10.1063/1.2149314}}.

\bibitem{luchini2009phase}
P.~Luchini and F.~Charru.
\newblock The phase lead of shear stress in shallow-water flow over a perturbed
  bottom.
\newblock {\em Bull. Am. Phys. Soc.}, 62:GS--005, 2009.

\bibitem{luchini2010phase}
P.~Luchini and F.~Charru.
\newblock The phase lead of shear stress in shallow-water flow over a perturbed
  bottom.
\newblock {\em Journal of fluid mechanics}, 665:516, 2010.
\newblock \href {https://doi.org/10.1017/S0022112010004313}
  {\path{doi:10.1017/S0022112010004313}}.

\bibitem{luchini2017quasilaminar}
P.~Luchini and F.~Charru.
\newblock Quasilaminar regime in the linear response of a turbulent flow to
  wall waviness.
\newblock {\em Physical Review Fluids}, 2(1):012601, 2017.
\newblock \href {https://doi.org/10.1103/PhysRevFluids.2.012601}
  {\path{doi:10.1103/PhysRevFluids.2.012601}}.

\bibitem{luchini2019fresh}
P.~Luchini and F.~Charru.
\newblock A fresh look at an old problem: perturbed flow over uneven terrain.
\newblock {\em Bull. Am. Phys. Soc.}, pages G18--001, 2019.

\bibitem{luchini2019large}
P.~Luchini and F.~Charru.
\newblock On the large difference between {Benjamin}'s and {Hanratty}'s
  formulations of perturbed flow over uneven terrain.
\newblock {\em Journal of Fluid Mechanics}, 871:534--561, 2019.
\newblock \href {https://doi.org/10.1017/jfm.2019.312}
  {\path{doi:10.1017/jfm.2019.312}}.

\bibitem{luchini2014error}
P.~Luchini and F.~Giannetti.
\newblock Error sensitivity to refinement: a criterion for optimal grid
  adaptation.
\newblock In {\em GIMC-GMA 2014 - XX Convegno Nazionale di Meccanica
  Computazionale VII - Riunione del Gruppo Materiali AIMETA}, pages 3--4.
  Universit\`a degli studi di Cassino e del Lazio Meridionale, 2014.

\bibitem{luchini2013shortAPS}
P.~Luchini, F.~Giannetti, and V.~Citro.
\newblock Short-wave analysis of {3D} and {2D} instabilities in a driven
  cavity.
\newblock {\em Bull. Am. Phys. Soc.}, pages L10--007, 2013.

\bibitem{luchini2013short}
P.~Luchini, F.~Giannetti, and V.~Citro.
\newblock Short-wave analysis of instabilities in open and closed cavities.
\newblock In {\em Euromech Colloquium 547 - Trends in open shear flow
  instability}, page~31. LadHyX, {\'E}cole polytechnique, 2013.

\bibitem{luchini2017error}
P.~Luchini, F.~Giannetti, and V.~Citro.
\newblock Error sensitivity to refinement: a criterion for optimal grid
  adaptation.
\newblock {\em Theoretical and Computational Fluid Dynamics}, 31(5-6):595--605,
  2017.
\newblock \href {https://doi.org/10.1007/s00162-016-0413-x}
  {\path{doi:10.1007/s00162-016-0413-x}}.

\bibitem{luchini2007structural}
P.~Luchini, F.~Giannetti, and J.~Pralits.
\newblock Structural sensitivity of the finite-amplitude vortex shedding behind
  a circular cylinder.
\newblock {\em Bull. Am. Phys. Soc.}, 60:BG--006, 2007.

\bibitem{luchini2008structural}
P.~Luchini, F.~Giannetti, and J.~Pralits.
\newblock Structural sensitivity of linear and nonlinear global modes.
\newblock In {\em 5th AIAA Theoretical Fluid Mechanics Conference}, page 4227,
  2008.
\newblock \href {https://doi.org/10.2514/6.2008-4227}
  {\path{doi:10.2514/6.2008-4227}}.

\bibitem{luchini2009structural}
P.~Luchini, F.~Giannetti, and J.~Pralits.
\newblock Structural sensitivity of the finite-amplitude vortex shedding behind
  a circular cylinder.
\newblock In {\em IUTAM Symposium on Unsteady Separated Flows and their
  Control}, pages 151--160. Springer, Dordrecht, 2009.
\newblock \href {https://doi.org/10.1007/978-1-4020-9898-7\_12}
  {\path{doi:10.1007/978-1-4020-9898-7\_12}}.

\bibitem{luchini2010optimal}
P.~Luchini, F.~Giannetti, and J.~Pralits.
\newblock Optimal control of a thin-airfoil wake using a {Riccati}-less
  approach.
\newblock In {\em Programme and Proceedings of 8th Euromech Fluid Mechanics
  Conference}, pages 13--16, 2010.

\bibitem{luchini2006stabilita}
P.~Luchini and A.~Pozzi.
\newblock Stabilita del flusso potenziale bidimensionale in prossimita del
  bordo di una superficie libera.
\newblock In {\em AIMETA2005: Atti del XVI Congresso AIMETA di Meccanica
  Teorica e Applicata, Firenze 11-15 Sep. 2005}, pages 1--11. Firenze
  University Press, 2006.

\bibitem{luchini2000directSIMAI}
P.~Luchini and M.~Quadrio.
\newblock Direct numerical simulation of turbulent flow in an annular pipe.
\newblock In {\em Minisimposio su ``Transizione e Turbolenza'' del V Congresso
  Nazionale della SIMAI}, pages 626--629, 2000.

\bibitem{luchini2000direct}
P.~Luchini and M.~Quadrio.
\newblock Direct simulation of turbulent flow in a pipe with annular
  cross-section.
\newblock In {\em 4th EUROMECH Fluid Mechanics Conference}, pages 33--33, 2000.

\bibitem{luchini20014}
P.~Luchini and M.~Quadrio.
\newblock A 4-th order accurate, parallel numerical method for the direct
  numerical simulation of turbulence in rectangular and cylindrical geometries.
\newblock In {\em XV Congresso Nazionale dell'Associazione Italiana di
  Meccanica Teorica e Applicata (AIMETA)}, pages 1--15, 2001.

\bibitem{luchini2001convection}
P.~Luchini and M.~Quadrio.
\newblock Convection velocity in turbulent wall flows.
\newblock In {\em XVI Congresso Nazionale AIDAA}, pages 1--10, 2001.

\bibitem{luchini2002adjoint}
P.~Luchini and M.~Quadrio.
\newblock Adjoint {DNS} of turbulent channel flow.
\newblock In {\em ASME 2002 Joint US-European Fluids Engineering Division
  Conference}, pages 1381--1385. American Society of Mechanical Engineers
  Digital Collection, 2002.
\newblock \href {https://doi.org/10.1115/FEDSM2002-31048}
  {\path{doi:10.1115/FEDSM2002-31048}}.

\bibitem{luchini2006low}
P.~Luchini and M.~Quadrio.
\newblock A low-cost parallel implementation of direct numerical simulation of
  wall turbulence.
\newblock {\em Journal of Computational Physics}, 211(2):551--571, 2006.
\newblock \href {https://doi.org/10.1016/j.jcp.2005.06.003}
  {\path{doi:10.1016/j.jcp.2005.06.003}}.

\bibitem{luchini2019model}
P.~Luchini and M.~Quadrio.
\newblock A model for fluctuations of the spatial mean in a turbulent channel
  flow.
\newblock In {\em European Drag Reduction and Flow Control Meeting, EDRFCM
  2019}, 2019.

\bibitem{luchini2006phase}
P.~Luchini, M.~Quadrio, and S.~Zuccher.
\newblock The phase-locked mean impulse response of a turbulent channel flow.
\newblock {\em Physics of Fluids}, 18(12):121702, 2006.
\newblock \href {https://doi.org/10.1063/1.2409729}
  {\path{doi:10.1063/1.2409729}}.

\bibitem{luchini2011comparisonAPS}
P.~Luchini and S.~Russo.
\newblock A comparison between eddy-viscosity models and direct numerical
  simulation: the response of turbulent flow to a volume force.
\newblock {\em Bulletin of the American Physical Society}, 56(18):41--41, 2011.

\bibitem{luchini2011comparison}
P.~Luchini and S.~Russo.
\newblock A comparison between eddy-viscosity models and direct numerical
  simulation: the response of turbulent flow to volume forcing.
\newblock In {\em XX Congresso AIMETA di Meccanica Teorica e Applicata, Bologna
  12-15 Sep. 2011}, pages 1--9, 2011.

\bibitem{luchini2015fast}
P.~Luchini and S.~Russo.
\newblock A fast algorithm for the estimation of statistical error in {DNS} (or
  experimental) time averages.
\newblock {\em Bull. Am. Phys. Soc.}, pages R5--007, 2015.

\bibitem{luchini1996direction}
P.~Luchini and R.~Tognaccini.
\newblock Direction-adaptive nonreflecting boundary conditions.
\newblock {\em Journal of Computational Physics}, 128(1):121--133, 1996.
\newblock \href {https://doi.org/10.1006/jcph.1996.0199}
  {\path{doi:10.1006/jcph.1996.0199}}.

\bibitem{luchini1999comparison}
P.~Luchini and R.~Tognaccini.
\newblock Comparison of viscous and inviscid numerical simulations of the
  start-up vortex issuing from a semi-infinite flat plate.
\newblock In {\em ESAIM: Proceedings}, volume~7, pages 247--257. EDP Sciences,
  1999.

\bibitem{luchini2002start}
P.~Luchini and R.~Tognaccini.
\newblock The start-up vortex issuing from a semi-infinite flat plate.
\newblock {\em Journal of Fluid Mechanics}, 455:175--193, 2002.
\newblock \href {https://doi.org/10.1017/S0022112001007340}
  {\path{doi:10.1017/S0022112001007340}}.

\bibitem{luchini2017viscous}
P.~Luchini and R.~Tognaccini.
\newblock Viscous and inviscid simulations of the start-up vortex.
\newblock {\em Journal of Fluid Mechanics}, 813:53--69, 2017.
\newblock \href {https://doi.org/10.1017/jfm.2016.867}
  {\path{doi:10.1017/jfm.2016.867}}.

\bibitem{marino2009adjoint}
L.~Marino and P.~Luchini.
\newblock Adjoint analysis of the flow over a forward-facing step.
\newblock {\em Theoretical and Computational Fluid Dynamics}, 23(1):37--54,
  2009.
\newblock \href {https://doi.org/10.1007/s00162-008-0090-5}
  {\path{doi:10.1007/s00162-008-0090-5}}.

\bibitem{martinelli2007active}
F.~Martinelli, M.~Quadrio, and P.~Luchini.
\newblock Active control and drag reduction in turbulent wall flows.
\newblock {\em Convegno Calcolo ad Alte Prestazioni. Milano}, 2007.

\bibitem{martinelli2007reynolds}
F.~Martinelli, M.~Quadrio, and P.~Luchini.
\newblock Reynolds-number dependence of the feedback control of turbulent
  channel flow.
\newblock In {\em XIX Congresso Nazionale AIDAA, Forli 17-21 Sep 2007}, pages
  1--11, 2007.

\bibitem{martinelli2010turbulent}
F.~Martinelli, M.~Quadrio, and P.~Luchini.
\newblock Turbulent drag reduction by feedback: a {Wiener}-filtering approach.
\newblock In {\em Advances in Turbulence XII: Proceedings of the 12th EUROMECH
  European Turbulence Conference, September 7-10, 2009, Marburg, Germany},
  pages 241--246. Springer, 2009.
\newblock \href {https://doi.org/10.1007/978-3-642-03085-7-59}
  {\path{doi:10.1007/978-3-642-03085-7-59}}.

\bibitem{martinelli2009wiener}
F.~Martinelli, M.~Quadrio, and P.~Luchini.
\newblock Wiener-hopf design of feedback compensators for drag reduction in
  turbulent channels.
\newblock In {\em XX Congresso Nazionale AIDAA - Milano, 2009}, 2009.

\bibitem{pirro.quadrio-2007-Directnumerical}
D.~Pirr{\`o} and M.~Quadrio.
\newblock Direct numerical simulation of turbulent
  {{Taylor}}\textendash{{Couette}} flow.
\newblock {\em Eur. J. Mech. B/Fluids}, doi: 10.1016/j.euromechflu.2007.10.005,
  2007.
\newblock \href {https://doi.org/10.1016/j.euromechflu.2007.10.005}
  {\path{doi:10.1016/j.euromechflu.2007.10.005}}.

\bibitem{Pralits2017}
J.~Pralits, E.~Alinovi, and A.~Bottaro.
\newblock Stability of the flow in a plane microchannel with one or two
  superhydrophobic walls.
\newblock {\em Physical Review Fluids}, 2(1), 2017.
\newblock \href {https://doi.org/10.1103/PhysRevFluids.2.013901}
  {\path{doi:10.1103/PhysRevFluids.2.013901}}.

\bibitem{pralits2008feedback}
J.~Pralits, T.~Bewley, and P.~Luchini.
\newblock Feedback stabilization of the wake behind a steady cylinder.
\newblock In {\em 7th ERCOFTAC SIG 33-FLUBIO WORKSHOP on Open Issues in
  Transition and Flow Control}, 2008.

\bibitem{Pralits2010513}
J.~Pralits, L.~Brandt, and F.~Giannetti.
\newblock Instability and sensitivity of the flow around a rotating circular
  cylinder.
\newblock {\em Journal of Fluid Mechanics}, 650:513--536, 2010.
\newblock \href {https://doi.org/10.1017/S0022112009993764}
  {\path{doi:10.1017/S0022112009993764}}.

\bibitem{Pralits20135}
J.~Pralits, F.~Giannetti, and L.~Brandt.
\newblock Three-dimensional instability of the flow around a rotating circular
  cylinder.
\newblock {\em Journal of Fluid Mechanics}, 730:5--18, 2013.
\newblock \href {https://doi.org/10.1017/jfm.2013.334}
  {\path{doi:10.1017/jfm.2013.334}}.

\bibitem{pralits2005leakyAPS}
J.~Pralits and P.~Luchini.
\newblock Leaky waves in boundary layer flow.
\newblock In {\em APS Division of Fluid Dynamics Meeting Abstracts}, volume~1,
  pages BAPS--2005, 2005.

\bibitem{pralits2009global}
J.~O. Pralits, F.~Giannetti, and P.~Luchini.
\newblock A global stability analysis of a thin-airfoil wake.
\newblock In {\em Atti del XIX Congresso AIMETA di Meccanica Teorica e
  Applicata Ancona (An), Italia 14-17 Settembre 2009}, pages 734--744. FANO
  ARAS EDIZIONI, 2009.

\bibitem{pralits2005leaky}
J.~O. Pralits and P.~Luchini.
\newblock Leaky waves in spatial stability analysis.
\newblock In {\em XVII Congresso AIMeTA di Meccanica Teorica e Applicata},
  pages 244--248. Firenze University Press, 2005.

\bibitem{pralits2010riccati}
J.~O. Pralits and P.~Luchini.
\newblock Riccati-less optimal control of bluff-body wakes.
\newblock In {\em Seventh IUTAM Symposium on Laminar-Turbulent Transition},
  pages 325--330. Springer, Dordrecht, 2010.
\newblock \href {https://doi.org/10.1007/978-90-481-3723-7-52}
  {\path{doi:10.1007/978-90-481-3723-7-52}}.

\bibitem{quadrio2003control}
M.~Quadrio, J.~Floryan, and P.~Luchini.
\newblock Control of turbulent channel flow using distributed suction.
\newblock In {\em 5th EUROMECH Fluid Mechanics Conference}, 2003.

\bibitem{quadrio2003modification}
M.~Quadrio, J.~Floryan, and P.~Luchini.
\newblock Modification of turbulent flow using distributed suction.
\newblock In {\em 50th Annual meeting of the Canadian Aeronautics and Space
  Institute}, pages 1--10, 2003.

\bibitem{quadrio2005modification}
M.~Quadrio, J.~Floryan, and P.~Luchini.
\newblock Modification of turbulent flow using distributed transpiration.
\newblock {\em Canadian Aeronautics and Space Journal}, 51(2):61--69, 2005.
\newblock \href {https://doi.org/10.5589/q05-008} {\path{doi:10.5589/q05-008}}.

\bibitem{quadrio2007effect}
M.~Quadrio, J.~Floryan, and P.~Luchini.
\newblock Effect of streamwise-periodic wall transpiration on turbulent
  friction drag.
\newblock {\em Journal of Fluid Mechanics}, 576(004):425--444, 2007.
\newblock \href {https://doi.org/10.1017/S0022112007004727}
  {\path{doi:10.1017/S0022112007004727}}.

\bibitem{quadrio.etal-2016-Doeschoice}
M.~Quadrio, B.~Frohnapfel, and Y.~Hasegawa.
\newblock Does the choice of the forcing term affect flow statistics in {{DNS}}
  of turbulent channel flow?
\newblock {\em Eur. J. Mech. B / Fluids}, 55:286--293, 2016.
\newblock \href {https://doi.org/10.1016/j.euromechflu.2015.09.005}
  {\path{doi:10.1016/j.euromechflu.2015.09.005}}.

\bibitem{quadrio2002direct}
M.~Quadrio and P.~Luchini.
\newblock Direct numerical simulation of the turbulent flow in a pipe with
  annular cross section.
\newblock {\em European Journal of Mechanics-B/Fluids}, 21(4):413--427, 2002.
\newblock \href {https://doi.org/10.1016/S0997-7546(02)01192-5}
  {\path{doi:10.1016/S0997-7546(02)01192-5}}.

\bibitem{quadrio2002linear}
M.~Quadrio and P.~Luchini.
\newblock The linear response of a turbulent channel flow.
\newblock In {\em 9th Euromech European Turbulence Conference (EETC9)}, pages
  715--718. CIMNE, 2002.

\bibitem{quadrio2003integral}
M.~Quadrio and P.~Luchini.
\newblock Integral space--time scales in turbulent wall flows.
\newblock {\em Physics of fluids}, 15(8):2219--2227, 2003.
\newblock \href {https://doi.org/10.1063/1.1586273}
  {\path{doi:10.1063/1.1586273}}.

\bibitem{quadrio2004numerical}
M.~Quadrio and P.~Luchini.
\newblock The numerical solution of the incompressible {Navier--Stokes}
  equations on a low cost, dedicated parallel computer.
\newblock {\em Preprint}, 2004.

\bibitem{quadrio2003parallel}
M.~Quadrio, P.~Luchini, and J.~Floryan.
\newblock A parallel algorithm for the direct numerical simulation of turbulent
  channel flow.
\newblock In {\em Proc. of the XI Conf. of the CFD Society of Canada}, pages
  28--30, 2003.

\bibitem{quadrio.ricco-2003-Initialresponse}
M.~Quadrio and P.~Ricco.
\newblock Initial response of a turbulent channel flow to spanwise oscillation
  of the walls.
\newblock {\em Journal of Turbulence}, 4(7), 2003.
\newblock \href {https://doi.org/10.1088/1468-5248/4/1/007}
  {\path{doi:10.1088/1468-5248/4/1/007}}.

\bibitem{quadrio.ricco-2004-Criticalassessment}
M.~Quadrio and P.~Ricco.
\newblock Critical assessment of turbulent drag reduction through spanwise wall
  oscillation.
\newblock {\em Journal of Fluid Mechanics}, 521:251--271, 2004.
\newblock \href {https://doi.org/10.1017/S0022112004001855}
  {\path{doi:10.1017/S0022112004001855}}.

\bibitem{quadrio.ricco-2011-laminargeneralized}
M.~Quadrio and P.~Ricco.
\newblock The laminar generalized {{Stokes}} layer and turbulent drag
  reduction.
\newblock {\em Journal of Fluid Mechanics}, 667:135--157, 2011.
\newblock \href {https://doi.org/10.1017/S0022112010004398}
  {\path{doi:10.1017/S0022112010004398}}.

\bibitem{quadrio.etal-2009-Streamwisetravelingwaves}
M.~Quadrio, P.~Ricco, and C.~Viotti.
\newblock Streamwise-traveling waves of spanwise wall velocity for turbulent
  drag reduction.
\newblock {\em Journal of Fluid Mechanics}, 627:161--178, 2009.
\newblock \href {https://doi.org/10.1017/S0022112009006077}
  {\path{doi:10.1017/S0022112009006077}}.

\bibitem{quadrio2007skin}
M.~Quadrio, C.~Viotti, and P.~Luchini.
\newblock Skin-friction drag reduction via steady streamwise oscillations of
  spanwise velocity.
\newblock In {\em Advances in Turbulence XI}, pages 659--661. Springer, Berlin,
  Heidelberg, 2007.
\newblock \href {https://doi.org/10.1007/978-3-540-72604-3\_210}
  {\path{doi:10.1007/978-3-540-72604-3\_210}}.

\bibitem{ricco.etal-2012-Changesturbulent}
P.~Ricco, C.~Ottonelli, Y.~Hasegawa, and M.~Quadrio.
\newblock Changes in turbulent dissipation in a channel flow with oscillating
  walls.
\newblock {\em Journal of Fluid Mechanics}, 700:77--104, 2012.
\newblock \href {https://doi.org/10.1017/jfm.2012.97}
  {\path{doi:10.1017/jfm.2012.97}}.

\bibitem{ricco.quadrio-2008-Walloscillationconditions}
P.~Ricco and M.~Quadrio.
\newblock Wall-oscillation conditions for drag reduction in turbulent channel
  flow.
\newblock {\em International Journal of Heat and Fluid Flow}, 29:601--612,
  2008.
\newblock \href {https://doi.org/10.1016/j.ijheatfluidflow.2007.12.005}
  {\path{doi:10.1016/j.ijheatfluidflow.2007.12.005}}.

\bibitem{rosti_brandt_pinelli_2018}
M.~E. Rosti, L.~Brandt, and A.~Pinelli.
\newblock Turbulent channel flow over an anisotropic porous wall drag increase
  and reduction.
\newblock {\em Journal of Fluid Mechanics}, 842:381--394, 2018.
\newblock \href {https://doi.org/10.1017/jfm.2018.152}
  {\path{doi:10.1017/jfm.2018.152}}.

\bibitem{rosti2014direct}
M.~E. Rosti, L.~Cortelezzi, and M.~Quadrio.
\newblock Direct numerical simulation of turbulent channel flow over porous
  walls.
\newblock {\em J Fluid Mech}, 2015.
\newblock \href {https://doi.org/doi:10.1017/jfm.2015.566}
  {\path{doi:doi:10.1017/jfm.2015.566}}.

\bibitem{russo2013linear}
S.~Russo and P.~Luchini.
\newblock The linear response of turbulent flow to an undulated wall.
\newblock In {\em XXI Congresso dell'Associazione Italiana di Meccanica Teorica
  ed Applicata (AIMETA)}, 2013.

\bibitem{russo2016linear}
S.~Russo and P.~Luchini.
\newblock The linear response of turbulent flow to a volume force: comparison
  between eddy-viscosity model and {DNS}.
\newblock {\em Journal of Fluid Mechanics}, 790:104--127, 2016.
\newblock \href {https://doi.org/10.1017/jfm.2016.4}
  {\path{doi:10.1017/jfm.2016.4}}.

\bibitem{russo2017fast}
S.~Russo and P.~Luchini.
\newblock A fast algorithm for the estimation of statistical error in {DNS} (or
  experimental) time averages.
\newblock {\em Journal of Computational Physics}, 347:328--340, 2017.
\newblock \href {https://doi.org/10.1016/j.jcp.2017.07.005}
  {\path{doi:10.1016/j.jcp.2017.07.005}}.

\bibitem{straub2017turbulent}
S.~Straub, R.~Vinuesa, P.~Schlatter, B.~Frohnapfel, and D.~Gatti.
\newblock Turbulent duct flow controlled with spanwise wall oscillations.
\newblock {\em Flow, Turbulence and Combustion}, 99(3-4):787--806, 2017.
\newblock \href {https://doi.org/10.1007/s10494-017-9846-6}
  {\path{doi:10.1007/s10494-017-9846-6}}.

\bibitem{viotti2009streamwise}
C.~Viotti, M.~Quadrio, and P.~Luchini.
\newblock Streamwise oscillation of spanwise velocity at the wall of a channel
  for turbulent drag reduction.
\newblock {\em Physics of fluids}, 21(11):115109, 2009.
\newblock \href {https://doi.org/10.1063/1.3266945}
  {\path{doi:10.1063/1.3266945}}.

\bibitem{zuccher2006algebraic}
S.~Zuccher, A.~Bottaro, and P.~Luchini.
\newblock Algebraic growth in a {Blasius} boundary layer: Nonlinear optimal
  disturbances.
\newblock {\em European Journal of Mechanics-B/Fluids}, 25(1):1--17, 2006.
\newblock \href {https://doi.org/10.1016/j.euromechflu.2005.07.001}
  {\path{doi:10.1016/j.euromechflu.2005.07.001}}.

\bibitem{zuccher2002time}
S.~Zuccher and P.~Luchini.
\newblock Time-dependent optimal perturbations for the algebraic instability in
  the nonlinear regime.
\newblock In {\em Fluids Engineering Division Summer Meeting}, volume 36150,
  pages 1387--1393, 2002.
\newblock \href {https://doi.org/10.1115/FEDSM2002-31049}
  {\path{doi:10.1115/FEDSM2002-31049}}.

\bibitem{zuccher2014boundary}
S.~Zuccher and P.~Luchini.
\newblock Boundary-layer receptivity to external disturbances using multiple
  scales.
\newblock {\em Meccanica}, 49(2):441--467, 2014.
\newblock \href {https://doi.org/10.1007/s11012-013-9804-x}
  {\path{doi:10.1007/s11012-013-9804-x}}.

\bibitem{zuccher2004algebraic}
S.~Zuccher, P.~Luchini, and A.~Bottaro.
\newblock Algebraic growth in a {Blasius} boundary layer: optimal and robust
  control by mean suction in the nonlinear regime.
\newblock {\em Journal of Fluid Mechanics}, 513:135, 2004.
\newblock \href {https://doi.org/10.1017/S0022112004000011}
  {\path{doi:10.1017/S0022112004000011}}.

\end{thebibliography}
\bibliographystyle{abbrvurl}
\addcontentsline{toc}{section}{References}
\end{document}